\documentclass{bioinfoArXiv}
\copyrightyear{2015}
\pubyear{2015}
\usepackage{verbatim} 
\usepackage{amssymb}
\usepackage{amsmath}
\usepackage{algorithm2e}
\usepackage{verbatim} 
\usepackage{epstopdf}
\usepackage{setspace}
\usepackage{amsthm}
\usepackage{rotating}
\usepackage[T1]{fontenc}
\usepackage{subfig}
\newtheorem{definition}{Definition}[section]%definition labelled with section

\newtheorem{theorem}{Theorem}
\newtheorem{thm}{Theorem}[section]

\newtheorem{prop}[thm]{Proposition}

\begin{document}
\firstpage{1}

\title[Periodic power spectrum and periodicities in DNA sequences]{Periodic power spectrum with applications in detection of latent periodicities in DNA sequences}
\author[Yin and Wang]{Changchuan Yin\,$^{1}$ \footnote{to whom correspondence should be addressed}, Jiasong Wang\,$^{2}$}
\address{$^{1}$Department of Mathematics, Statistics and Computer Science\\ The University of Illinois at Chicago, Chicago, IL 60607-7045, USA\\
$^{2}$Department of Mathematics, Nanjing University, Nanjing 210093, China.}

\history{} %Received on XXXXX; revised on XXXXX; accepted on XXXXX}

\editor{} %Associate Editor: XXXXXXX}

\maketitle

\begin{abstract}

\section{Motivation:}
Latent periodic elements in genomes play important roles in genomic functions. Many complex periodic elements in genomes are difficult to be detected by commonly used digital signal processing (DSP). We present a novel method to compute the periodic power spectrum of a DNA sequence based on the nucleotide distributions on periodic positions of the sequence. The method directly calculates full periodic spectrum of a DNA sequence rather than frequency spectrum by Fourier transform. The magnitude of the periodic power spectrum reflects the strength of the periodicity signals, thus, the algorithm can capture all the latent periodicities in DNA sequences.
\section{Results:}
We apply this method on detection of latent periodicities in different genome elements, including exons and microsatellite DNA sequences. The results show that the method minimizes the impact of spectral leakage, captures a much broader latent periodicities in genomes, and outperforms the conventional Fourier transform.
\section{Availability:}
The source code for the algorithms described in this paper can be accessed at MATLAB file central.

\section{Contact:} \href{cyin1@uic.edu}{cyin1@uic.edu}
\end{abstract}

\section{Introduction}
\label{Introduction}
%YIN0204:Functions of periodicities
Regions of approximate tandem repeats in genomes are abundant in many species from bacteria to mammals, and are essential for many structures and functions of genomes \citep{shapiro2005repetitive}. For example, many researchers revealed that the 3-periodicity of a DNA sequence indicates protein coding regions and it is used for protein coding region identication \citep{silverman1986measure,tiwari1997prediction}. Pervasive hidden $10-11 bp$ periodicities in complete genomes reflect protein structure and DNA folding \citep{herzel199910}. In bacteria genomes, repetitive DNA sequences form alternative DNA conformations and thus induce genetic instability \citep{wojcik2012direct}. Repetitive DNA sequence elements are also fundamental to the cooperative molecular interactions forming nucleoprotein complexes. In a human genome, simple tandem DNA repeats may associate with human neurodegenerative diseases \citep{sutherland1995simple}. The short tandem repeats (STRs) may have a wide range of applications, including medical genetics, forensics, and genetic genealogy \citep{gymrek2012lobstr}. In addition, repeats in genomes often cause assembling and indels discovery problems in next-generation sequencing \citep{treangen2011repetitive,narzisi2015challenge}.

%YIN0204:Current methods and problems
Repetitive sequences and periodicities in genomes vary in size, scale and complexity \citep{trifonov19983,hauth2002beyond}. The size of repeats can be from two base pairs to hundreds of base pairs.  They can disperse widely in a relatively long sequence range, and can be just a tandem arrays of simple sequence composition. Some repeats contain partial periodic sequence patterns, and some have hidden periodicity due to multiple periodic components \citep{korotkov2003information}. Thus, latent repeat signals in DNA sequences are difficult to be analyzed by straightforward observation and comparison. Accurate identification these repeats and periodicities is one of important problems in genome analysis.
 
The latent periodicities and repetitive features of DNA sequences are primarily studied by digital signal processing (DSP) approaches such as Fourier transform \citep{silverman1986measure,sharma2004spectral,buchner2003detection}, and wavelet transform \citep{wang2010localizing}. The DNA sequences are converted to a numeric sequence and then Fourier transform is applied for power spectrum analysis in the DNA sequence \citep{afreixo2004fourier}. The other periodicity detection methods include statistics method \citep{epps2011statistical}, maximum likelihood estimation \citep{arora2007detection}, information decomposition \citep{korotkov2003information}, direct frequency mapping \citep{glunvcic2013direct}, and chaos game representation (CGR) \citep{messaoudi2014building}. For review of these methods, refer to \cite{grover2012searching}. In spite of many studies on hidden periodicities in DNA sequences, due to the complexity of repetitive sequences and large amount of background noise presented in Fourier analysis and other methods, identification of all the hidden periodicities in DNA sequences is still a challenging problem \citep{suvorova2014comparative,epps2011statistical,illingworth2008criteria}. 

%YIN0204: Our method and brief results
In this paper, we present a novel algorithm for computing periodic power spectrum (PPS) based on periodic distribution of nucleotides of DNA sequences. We apply the algorithm to the identification of hidden periodicities in DNA sequences. The results demonstrate that this algorithm is effective and efficient to capture quantitatively all the latent periodicities in DNA sequences and outperforms the methods based on Fourier transform.
 
\section{Methods and Algorithms}
 \subsection{Numerical representations of DNA sequences}
 DNA molecules consist of four linearly linked nucleotides, adenine (A), thymine (T), cytosine (C), and guanine (G). A DNA sequence can be represented as a permutation of four characters A, T, C, G of different lengths. To apply DSP in DNA sequence analysis, the character strings of DNA molecules must be mapped into one or more numerical sequences \citep{yin2014representation}. One of the methods in literatures is to use binary indicator sequences \citep{Voss1992}. A DNA sequence, denoted as, $s(0), s(1), \ldots, s(N-1)$, can be decomposed into four binary indicator sequences, $u_A(n), u_T(n), u_C(n)$, and $u_G(n)$, which indicate the presence or absence of four nucleotides, $A, T, C$, and $G$ at the $n-$th position, respectively. The indicator mapping of DNA sequences is defined as follows:
  \begin{equation}
 %$$
 u_\alpha  (n) = \left\{ \begin{gathered}
   1,s(n) = \alpha  \hfill \\
   0,otherwise \hfill \\
 \end{gathered}  \right.
 %$$
  \end{equation}
 where  $\alpha\in{A, C, G, T}, n = 0,1,2, \ldots ,N - 1 $. The four indicator sequences correspond to the appearance of the four nucleotides at each position of the DNA sequence. For example, the indicator sequence, $u_A(n) =0001010111\ldots$, indicates that the nucleotide $A$ presents in the positions of 4, 6, 8, 9, and 10 of the DNA sequence. Table 1 illustrates and example of the Voss 4D binary indicator mappings of a DNA sequence.
 
\begin{table}[ht]
 \caption{The Voss 4D binary indicator mappings of a short DNA sequence.}
 \centering % used for centering table
 %\begin{center}
 \begin{tabular}{l*{12}{c}r}
 \hline\hline
 DNA   & T & A & G & C & C & T & G & C & T & G & A & T\\
 \hline
 $u_A$  & 0 & 1 & 0 & 0 & 0 & 0 & 0 & 0 & 0 & 0 & 1 & 0\\
 $u_T$  & 1 & 0 & 0 & 0 & 0 & 1 & 0 & 0 & 1 & 0 & 0 & 1\\
 $u_C$  & 0 & 0 & 0 & 1 & 1 & 0 & 0 & 1 & 0& 0 & 0 & 0 \\
 $u_G$  & 0 & 0  & 1  & 0  &  0 & 0 & 1 & 0& 0 & 1 & 0&  0\\
 \hline\hline
 \end{tabular}
 %\end{center}
 \label{table:nonlin} % is used to refer this table in the text
 \end{table}

 \subsection{Fourier power spectrum}
 Discrete Fourier transform (DFT) is the transformation of observation data in time domain to new values in frequency domain. It gives a unique representation of the original signal in frequency domain. DFT spectral analysis of DNA sequences may detect any latent or hidden periodical signal in the original sequences. It may discover approximate repeats that are difficult to be detected by tandem repeat search. Let $X$ be the DFT of real number time series $x$ of length $N$, then $X(k)$ is defined as\\
 \begin{equation}\label{e:WeakBase}
  % MathType!MTEF!2!1!+-
  % feaafiart1ev1aaatCvAUfeBSjuyZL2yd9gzLbvyNv2CaerbuLwBLn
  % hiov2DGi1BTfMBaeXatLxBI9gBaerbd9wDYLwzYbItLDharqqtubsr
  % 4rNCHbWexLMBbXgBd9gzLbvyNv2CaeHbl7mZLdGeaGqiVCI8FfYJH8
  % YrFfeuY-Hhbbf9v8qqaqFr0xc9pk0xbba9q8WqFfeaY-biLkVcLq-J
  % Hqpepeea0-as0Fb9pgeaYRXxe9vr0-vr0-vqpWqaaeaabiGaciaaca
  % qabeaadaqaaqaafaGcbaGaamiwaiaacIcacaWGRbGaaiykaiabg2da
  % 9maaqahabaGaamiEaiaacIcacaWGUbGaaiykaiaadwgadaahaaWcbe
  % qaaiabgkHiTiaadMgacaaIYaGaeqiWda3aaSaaaeaacaWGRbaabaGa
  % amOtaaaacaWGUbaaaaqaaiaad6gacqGH9aqpcaaIWaaabaGaamOtai
  % abgkHiTiaaigdaa0GaeyyeIuoakiaacYcacaWGRbGaeyypa0JaaGim
  % aiaacYcacaaIXaGaaiilaiabl+UimjaacYcacaWGobGaeyOeI0IaaG
  % ymaaaa!60B6!
  %\[
  X(k) = \sum\limits_{n = 0}^{N - 1} {x(n)e^{ - i2\pi \frac{k}
  {N}n} } ,k = 0,1, \cdots ,N - 1
  %\]
 \end{equation}
 where $i = \sqrt{-1}$. The frequency domain vector $X$ contains all the information about the original signal $x$ and can recover the signal. The DFT power spectrum of the signal
 $x(n)$ at the frequency $k$ is defined as
 \begin{equation}
% MathType!MTEF!2!1!+-
% feaafiart1ev1aaatCvAUfeBSjuyZL2yd9gzLbvyNv2CaerbuLwBLn
% hiov2DGi1BTfMBaeXatLxBI9gBaerbd9wDYLwzYbItLDharqqtubsr
% 4rNCHbWexLMBbXgBd9gzLbvyNv2CaeHbl7mZLdGeaGqiVCI8FfYJH8
% YrFfeuY-Hhbbf9v8qqaqFr0xc9pk0xbba9q8WqFfeaY-biLkVcLq-J
% Hqpepeea0-as0Fb9pgeaYRXxe9vr0-vr0-vqpWqaaeaabiGaciaaca
% qabeaadaqaaqaafaGcbaGaamiuaiaadofacaGGOaGaam4AaiaacMca
% cqGH9aqpcaWGybGaaiikaiaadUgacaGGPaGaamiwaiaacIcacaWGRb
% GaaiykamaaCaaaleqabaGaaiOkaaaakiaacYcacaWGRbGaeyypa0Ja
% aGimaiaacYcacaaIXaGaaiOlaiabl+UimjaacYcacaWGobGaeyOeI0
% IaaGymaaaa!55D0!
%\[
PS(k) = X(k)X(k)^* ,k = 0,1. \cdots ,N - 1
%\]
\end{equation}
where $X(k)^*$ denotes the complex conjugate of $X(k)$. Similarly, for a DNA sequence, let $U_\alpha$ and $PS_\alpha$ be the Fourier transform and corresponding Fourier power spectrum of the binary indicator sequence $u_\alpha,\alpha  \in \{ A,T,C,G\}$ the DNA sequence, the Fourier power spectrum  of the DNA sequence is the sum of Fourier power spectra $PS_\alpha$. 
\begin{equation}
  % MathType!MTEF!2!1!+-
  % feaafiart1ev1aaatCvAUfeBSjuyZL2yd9gzLbvyNv2CaerbuLwBLn
  % hiov2DGi1BTfMBaeXatLxBI9gBaerbd9wDYLwzYbItLDharqqtubsr
  % 4rNCHbWexLMBbXgBd9gzLbvyNv2CaeHbl7mZLdGeaGqiVCI8FfYJH8
  % YrFfeuY-Hhbbf9v8qqaqFr0xc9pk0xbba9q8WqFfeaY-biLkVcLq-J
  % Hqpepeea0-as0Fb9pgeaYRXxe9vr0-vr0-vqpWqaaeaabiGaciaaca
  % qabeaadaqaaqaafaGcbaGaamiuaiaadofacaGGOaGaam4AaiaacMca
  % cqGH9aqpdaaeqbqaaiaadcfacaWGtbWaaSbaaSqaaiabeg7aHbqaba
  % GccaGGOaGaam4AaiaacMcaaSqaaiabeg7aHjabgIGiolaacUhacaWG
  % bbGaaiilaiaadsfacaGGSaGaam4qaiaacYcacaWGhbGaaiyFaaqab0
  % GaeyyeIuoaaaa!565D!
  %\[
  PS(k) = \sum\limits_{\alpha  \in \{ A,T,C,G\} } {PS_\alpha  (k)} 
  %\]
 \end{equation}
  
 The Fourier power spectrum of DNA sequences can be used to identify periodicities in DNA sequence. For example, most of protein coding regions show a prominent peak at frequency $k=N/3$ in Fourier power spectrum because of the variance of nucleotide distributions in the three codon positions \citep{yin2005fourier}. The Fourier power spectrum is large when the nucleotide has a significant tendency of appearing about every $p$ positions. In particular, when $k=N/3$, namely $\alpha$ tends to appear at a certain codon position. It shall be noted that because Fourier power spectra of real number series are symmetric, the Fourier power spectrum plots only show the first half of the spectrum. Furthermore, because DFT power spectrum at frequency zero is a constant and equivalent to the sum of data points, the power spectrum at frequency zero is not included in plotting. 

 \subsection{Periodic power spectrum}
The strength of a periodicity within a 1D real number signal is determined by the distribution of real values on periodic positions. Thus, the congruence derivative vector of a 1D real number signal, which represents these periodic distribution, was introduced by \cite{wang2012Some}. For DNA sequences, it has been shown that Fourier power spectrum at a periodicity $p$ is determined by the unique distribution of nucleotides on the periodic positions. For example, the 3-base periodicity is determined non-uniformed distribution of nucleotides on the three positions\citep{yin2005fourier}. Here, we define the congruent derivative vectors that reflect the nucleotide distributions on the periodic position. For example, for a DNA sequence, AGTTAACGCCTAGCC, when it is mapped into the Voss 4D binary sequences  $u_A,u_T,u_C,u_G$, of the congruence derivative vectors of periodicity 3 reflect the nucleotide distributions on periodic-3 positions in the DNA sequence, and have the following values: $f_{A }  = [1,1,2],f_{T }  = [1,1,1],f_{C }  = [2,1,2],f_{G }  = [1,2,0]$, respectively. 

The congruent derivative vector of a 1D real number signal is defined as follow
\begin{definition} For a real number signal $x$ of length $N$, the element of congruence derivative vector $f_p$ of the signal $x$, for periodicity $p$, is defined as\\
       \begin{equation}
 % MathType!MTEF!2!1!+-
% feaafiart1ev1aaatCvAUfeBSjuyZL2yd9gzLbvyNv2CaerbuLwBLn
% hiov2DGi1BTfMBaeXatLxBI9gBaerbd9wDYLwzYbItLDharqqtubsr
% 4rNCHbWexLMBbXgBd9gzLbvyNv2CaeHbl7mZLdGeaGqiVCI8FfYJH8
% YrFfeuY-Hhbbf9v8qqaqFr0xc9pk0xbba9q8WqFfeaY-biLkVcLq-J
% Hqpepeea0-as0Fb9pgeaYRXxe9vr0-vr0-vqpWqaaeaabiGaciaaca
% qabeaadaqaaqaafaGceaqabeaacaWGMbWaaSbaaSqaaiaadchaaeqa
% aOGaaiikaiaadghacaGGPaGaeyypa0ZaaabuaeaacaWG4bGaaiikai
% aad6gacaGGPaaaleaaciGGTbGaai4BaiaacsgacaGGOaGaamOBaiaa
% cYcacaWGWbGaaiykaiabg2da9iaadghaaeqaniabggHiLdaakeaaca
% WGXbGaeyypa0JaaGimaiaacYcacaaIXaGaaiilaiabl+UimjaacYca
% caWGWbGaeyOeI0IaaGymaiaacYcacaWGUbGaeyypa0JaaGimaiaacY
% cacaaIXaGaaiilaiabl+UimjaacYcacaWGobGaeyOeI0IaaGymaaaa
% aa!6753!
%\[
\begin{gathered}
  f_p (q) = \sum\limits_{\bmod (n,p) = q} {x(n)}  \hfill \\
  q = 0,1, \cdots ,p - 1,n = 0,1, \cdots ,N - 1 \hfill \\ 
\end{gathered} 
%\]
\end{equation}
  where $mod(n,p)$ is the modulo operation and returns the remainder after division of $n$ by $p$, then $f_p  = \left( {f_p (0),f_p (1), \cdots ,f_p (p - 1)} \right)$.
    \end{definition}
 We may consider a frequency domain transform is to project the 1D real number signal $x$ into some periodic shift sequences. Let $ \omega _p  = e^{ - i\frac{{2\pi }}{p}}$ be the $p-$th root of unity, we have the following periodic shift sequences as basis vectors. 
      \begin{equation}
      % MathType!MTEF!2!1!+-
      % feaafiart1ev1aqatCvAUfeBSjuyZL2yd9gzLbvyNv2CaerbuLwBLn
      % hiov2DGi1BTfMBaeXatLxBI9gBaerbd9wDYLwzYbItLDharqqtubsr
      % 4rNCHbWexLMBbXgBd9gzLbvyNv2CaeHbl7mZLdGeaGqiVCI8FfYJH8
      % YrFfeuY-Hhbbf9v8qqaqFr0xc9pk0xbba9q8WqFfeaY-biLkVcLq-J
      % Hqpepeea0-as0Fb9pgeaYRXxe9vr0-vr0-vqpWqaaeaabiGaciaaca
      % qabeaadaqaaqaafaGceaqabeaacqaH0oazdaqhaaWcbaGaamiCaaqa
      % aiaadghaaaGccaGGOaGaamOBaiaacMcacqGH9aqpdaGabaabaeqaba
      % GaeqyYdC3aa0baaSqaaiaadchaaeaacaWGXbaaaOGaaiilaiaabMga
      % caqGMbGaaeiiaiaab2gacaqGVbGaaeizaiaabIcacaqGUbGaaeylai
      % aabghacaqGSaGaaeiCaiaabMcacqGH9aqpcaaIWaaabaGaaGimaiaa
      % cYcacaqGVbGaaeiDaiaabIgacaqGLbGaaeOCaiaabEhacaqGPbGaae
      % 4CaiaabwgaaaGaay5EaaaabaGaaeilaiaabEhacaqGObGaaeyzaiaa
      % bkhacaqGLbGaaGjcVlaayIW7caaMi8UaamyCaiabg2da9iaaicdaca
      % GGSaGaaGymaiaacYcacqWIVlctcaGGSaGaamiCaiabgkHiTiaaigda
      % aaaa!7674!
      \begin{gathered}
        \delta _p^q (n) = \left\{ \begin{gathered}
        \omega _p^q ,{\text{if mod(n - q,p)}} = 0 \hfill \\
        0,{\text{otherwise}} \hfill \\ 
      \end{gathered}  \right. \hfill \\
        {\text{\ where}}{\kern 1pt} {\kern 1pt} {\kern 1pt} q = 0,1, \cdots ,p - 1 \hfill \\ 
      \end{gathered} 
        \end{equation}
  Table 2 is an example of the periodic shift sequences  $\delta _p^q $ as basis vector for periodicity 4.  
       \begin{table}[ht]
          \caption{Periodic shift sequences for periodicity 4} % title of Table
          \centering % used for centering table
          %\begin{center}
          \begin{tabular}{l*{12}{c}r}
          \hline\hline
          $n$  & 0 & 1 & 2 & 3 & 4 & 5 & 6 & 7 & $\cdots$ \\
          \hline
          $\delta _4^0 (n)$  & $\omega _4^0 $ & 0 & 0 & 0 & $\omega _4^0 $ & 0 & 0 & 0  & $\cdots$ \\
          $\delta _4^1 (n)$  & 0 & $\omega _4^1 $ & 0 & 0 & 0 & $\omega _4^1 $ & 0 & 0  & $\cdots$ \\
          $\delta _4^2 (n)$  & 0 & 0 & $\omega _4^2$ & 0 & 0 & 0 & $\omega _4^2 $ & 0  & $\cdots$ \\
          $\delta _4^3 (n)$  & 0 & 0 & 0 & $\omega _4^3 $ & 0 & 0 & 0 & $\omega _4^3 $  & $\cdots$ \\
          \hline\hline
          \end{tabular}
          %\end{center}
          \label{table:nonlin} % is used to refer this table in the text
       \end{table}
        
  We propose here the periodic transform of a real number signal $x$ at periodicity $p$ as the projection the signal onto the corresponding periodic basis vectors $\delta _p^q $ as in equation (7). The periodic transform is defined as
        \begin{definition}
    % MathType!MTEF!2!1!+-
    % feaafiart1ev1aaatCvAUfeBSjuyZL2yd9gzLbvyNv2CaerbuLwBLn
    % hiov2DGi1BTfMBaeXatLxBI9gBaerbd9wDYLwzYbItLDharqqtubsr
    % 4rNCHbWexLMBbXgBd9gzLbvyNv2CaeHbl7mZLdGeaGqiVCI8FfYJH8
    % YrFfeuY-Hhbbf9v8qqaqFr0xc9pk0xbba9q8WqFfeaY-biLkVcLq-J
    % Hqpepeea0-as0Fb9pgeaYRXxe9vr0-vr0-vqpWqaaeaabiGaciaaca
    % qabeaadaqaaqaafaGcbaGaamiwaiaadcfacaGGOaGaamiCaiaacMca
    % cqGH9aqpdaaeWbqaamaaqahabaGaamiEaiaacIcacaWGUbGaaiykai
    % abes7aKnaaDaaaleaacaWGWbaabaGaamyCaaaakiaacIcacaWGUbGa
    % aiykaaWcbaGaamyCaiabg2da9iaaicdaaeaacaWGWbGaeyOeI0IaaG
    % ymaaqdcqGHris5aaWcbaGaamOBaiabg2da9iaaicdaaeaacaWGobGa
    % eyOeI0IaaGymaaqdcqGHris5aOGaaiilaiaadchacqGH9aqpcaaIXa
    % GaaiilaiaaikdacaGGSaGaeS47IWKaaiilaiaad6eaaaa!6571!
    \[
    XP(p) = \sum\limits_{n = 0}^{N - 1} {\sum\limits_{q = 0}^{p - 1} {x(n)\delta _p^q (n)} } ,p = 1,2, \cdots ,N
    \]
     \end{definition}
  Specially, if the length of a signal is equivalent to a multiple of a periodicity $p$, we have following theorem. 
  \begin{theorem}
  If $N$ is equivalent to a multiple of $p$, the projection of signal $x$ to the periodic basis sequences $\delta _p^q $ is equivalent to Fourier transform of the congruent derivative vector of the signal.
  \end{theorem}
  The proof of the theorem 1 is provided by \cite{wang2012Some}. Let $\delta _p  = (\omega _p^0 ,\omega _p^1 , \cdots ,\omega _p^{p - 1} )$, the periodic transform at periodicity p may be re-written as
   \begin{equation}
 % MathType!MTEF!2!1!+-
 % feaafiart1ev1aaatCvAUfeBSjuyZL2yd9gzLbvyNv2CaerbuLwBLn
 % hiov2DGi1BTfMBaeXatLxBI9gBaerbd9wDYLwzYbItLDharqqtubsr
 % 4rNCHbWexLMBbXgBd9gzLbvyNv2CaeHbl7mZLdGeaGqiVCI8FfYJH8
 % YrFfeuY-Hhbbf9v8qqaqFr0xc9pk0xbba9q8WqFfeaY-biLkVcLq-J
 % Hqpepeea0-as0Fb9pgeaYRXxe9vr0-vr0-vqpWqaaeaabiGaciaaca
 % qabeaadaqaaqaafaGcbaGaamiwaiaadcfacaGGOaGaamiCaiaacMca
 % cqGH9aqpcaWGMbWaaSbaaSqaaiaadchaaeqaaOGaeqiTdq2aa0baaS
 % qaaiaadchaaeaacaWGubaaaaaa!49C9!
 %\[
 XP(p) = f_p \delta _p^T 
 %\]
 \end{equation}
 where $\delta_p^T$ is the transpose of $\delta_p$.
 
The power spectrum at periodicity $p$ can be then expressed as
 \begin{equation}
% MathType!MTEF!2!1!+-
% feaafiart1ev1aaatCvAUfeBSjuyZL2yd9gzLbvyNv2CaerbuLwBLn
% hiov2DGi1BTfMBaeXatLxBI9gBaerbd9wDYLwzYbItLDharqqtubsr
% 4rNCHbWexLMBbXgBd9gzLbvyNv2CaeHbl7mZLdGeaGqiVCI8FfYJH8
% YrFfeuY-Hhbbf9v8qqaqFr0xc9pk0xbba9q8WqFfeaY-biLkVcLq-J
% Hqpepeea0-as0Fb9pgeaYRXxe9vr0-vr0-vqpWqaaeaabiGaciaaca
% qabeaadaqaaqaafaGcbaGaamiuaiaadcfacaWGtbGaaiikaiaadcha
% caGGPaGaeyypa0JaamOzamaaBaaaleaacaWGWbaabeaakiaacIcacq
% aH0oazdaqhaaWcbaGaamiCaaqaaiaadsfaaaGccqaH0oazdaWgaaWc
% baGaamiCaaqabaGccaGGPaGaamOzamaaDaaaleaacaWGWbaabaGaam
% ivaaaaaaa!51B2!
%\[
PPS(p) = f_p (\delta _p^T \delta _p )f_p^T 
%\]
  \end{equation}
 where $f_p^T$ is the transpose of $f_p$ and $\delta _p^T$ is the transpose of $\delta _p$. The matrix that is formed by $\delta _p^T \delta _p$ only depends on periodicity $p$ and is a complex symmetric matrix with the main diagonal elements one. If we add the corresponding upper and lower entries of the matrix, we can get a lower-triangular real number matrix, $Sp$. The matrix $Sp$ can be used to compute the periodic power spectrum. The entries of the matrix $S_p$ are defined as:
   \begin{equation}
   % MathType!MTEF!2!1!+-
   % feaafiart1ev1aaatCvAUfeBSjuyZL2yd9gzLbvyNv2CaerbuLwBLn
   % hiov2DGi1BTfMBaeXatLxBI9gBaerbd9wDYLwzYbItLDharqqtubsr
   % 4rNCHbWexLMBbXgBd9gzLbvyNv2CaeHbl7mZLdGeaGqiVCI8FfYJH8
   % YrFfeuY-Hhbbf9v8qqaqFr0xc9pk0xbba9q8WqFfeaY-biLkVcLq-J
   % Hqpepeea0-as0Fb9pgeaYRXxe9vr0-vr0-vqpWqaaeaabiGaciaaca
   % qabeaadaqaaqaafaGcbaGaam4uamaaBaaaleaacaWGWbaabeaakiaa
   % cIcacaWGRbGaaiilaiaadQgacaGGPaGaeyypa0Zaaiqaaqaabeqaai
   % aaikdaciGGJbGaai4BaiaacohadaWcaaqaaiaaikdacqaHapaCcaGG
   % OaGaam4AaiabgkHiTiaaigdacaGGPaaabaGaamiCaaaaciGGJbGaai
   % 4BaiaacohadaWcaaqaaiaaikdacqaHapaCcaGGOaGaamOAaiabgkHi
   % TiaaigdacaGGPaaabaGaamiCaaaaaeaacaaMi8UaaGjcVlaayIW7ca
   % aMi8UaaGjcVlaayIW7caaMi8UaaGjcVlaayIW7caaMi8UaaGjcVlaa
   % yIW7caaMi8UaaGjcVlaayIW7caaMi8Uaey4kaSIaaGOmaiGacohaca
   % GGPbGaaiOBamaalaaabaGaaGOmaiabec8aWjaacIcacaWGRbGaeyOe
   % I0IaaGymaiaacMcaaeaacaWGWbaaaiGacohacaGGPbGaaiOBamaala
   % aabaGaaGOmaiabec8aWjaacIcacaWGQbGaeyOeI0IaaGymaiaacMca
   % aeaacaWGWbaaaiaacYcacaqGPbGaaGjcVlaabAgacaaMi8UaaGjcVl
   % aabUgacaqG+aGaaeOAaaqaaiaaigdacaGGSaGaaGjcVlaayIW7caqG
   % PbGaaeOzaiaayIW7caaMi8Uaae4Aaiaab2dacaqGQbaabaGaaGimai
   % aacYcacaqGPbGaaeOzaiaayIW7caaMi8Uaae4AaiaabYdacaqGQbaa
   % aiaawUhaaaaa!AA89!
   %\[
   S_p (k,j) = \left\{ \begin{gathered}
     2\cos \frac{{2\pi (k - 1)}}
   {p}\cos \frac{{2\pi (j - 1)}}
   {p} \hfill \\
     {\kern 1pt} {\kern 1pt} {\kern 1pt} {\kern 1pt} {\kern 1pt} {\kern 1pt} {\kern 1pt} {\kern 1pt} {\kern 1pt} {\kern 1pt} {\kern 1pt} {\kern 1pt} {\kern 1pt} {\kern 1pt} {\kern 1pt} {\kern 1pt}  + 2\sin \frac{{2\pi (k - 1)}}
   {p}\sin \frac{{2\pi (j - 1)}}
   {p},{\text{i}}{\kern 1pt} {\text{f}}{\kern 1pt} {\kern 1pt} {\text{k > j}} \hfill \\
     1,{\kern 1pt} {\kern 1pt} {\text{if}}{\kern 1pt} {\kern 1pt} {\text{k = j}} \hfill \\
     0,{\text{if}}{\kern 1pt} {\kern 1pt} {\text{k < j}} \hfill \\ 
   \end{gathered}  \right.
   %\]
   \end{equation}
 
 Based on above theories, we propose the following  algorithm (Algorithm $1$) to compute the power spectrum of a 1D real number signal using congruence derivative vector and the periodic shift sequences  $\delta _p^q $. Because the power spectrum from the algorithm corresponds to a specific periodicity, while a DFT power spectrum corresponds to a specific frequency, we name the method as PPS (Periodic Power Spectrum) algorithm.
  
   \begin{algorithm}
   \DontPrintSemicolon % Some LaTeX compilers require to use \dontprintsemicolon instead
   \KwIn{a real number signal $x$ of $N$, periodicity $p$}
   \KwOut{PPS at periodicity $p$}
    \textbf{Step:}
    \begin{enumerate}
     \item Generate a vector $ C = [1, cos (2 \pi / p), cos (4 \pi / p), ........, cos (2 (p-1)\pi /p)]$.
        \item Generate a vector $ V = [0, sin (2 \pi /p), sin (4 \pi / p), ........, sin (2 (p-1)\pi /p)]$.
        \item Obtain an matrix $U = C^{T}C +  V^{T}V$, where $C^{T}$ and $V^{T}$ are the transposes of vectors $C$ and $V$, respectively.
        \item Construct the spectrum transform matrix $S_p$ of size $p \times p$ from the matrix $U$:
             
             \lIf{$k>j$}
            {
                 \Return{$S_p(k,j)$=U(k,j)+U(j,k)}\;
            }
            \lIf{$k=j$}
            {
                 \Return{$S_p(k,j)$= 1}\;
            }
            
            \lElse{
            \Return{$S_p(k,j)$=0}\;
            }
       \item Compute the congruent derivative vector $f_p$ of the signal at periodicity $p$ (Equation (5)).
          \item Compute the periodic power spectrum $PPS(p)$ by $S_p$ and $f_p$:
          %\begin{equation}
          % MathType!MTEF!2!1!+-
          % feaafiart1ev1aaatCvAUfeBSjuyZL2yd9gzLbvyNv2CaerbuLwBLn
          % hiov2DGi1BTfMBaeXatLxBI9gBaerbd9wDYLwzYbItLDharqqtubsr
          % 4rNCHbWexLMBbXgBd9gzLbvyNv2CaeHbl7mZLdGeaGqiVCI8FfYJH8
          % YrFfeuY-Hhbbf9v8qqaqFr0xc9pk0xbba9q8WqFfeaY-biLkVcLq-J
          % Hqpepeea0-as0Fb9pgeaYRXxe9vr0-vr0-vqpWqaaeaabiGaciaaca
          % qabeaadaqaaqaafaGcbaGaamiuaiaadcfacaWGtbGaaiikaiaadcha
          % caGGPaGaeyypa0JaamOzamaaBaaaleaacaWGWbaabeaakiabgEHiQi
          % aadofadaWgaaWcbaGaamiCaaqabaGccqGHxiIkcaWGMbWaa0baaSqa
          % aiaadchaaeaacaWGubaaaaaa!4DC0!
          \[
          PPS(p) = f_p  * S_p  * f_p^T 
          \]
          % \end{equation}
           where ${f_p}^T$ is the transpose of congruent derivative vector $f_p$.
          \end{enumerate}
          \caption{Algorithm for computing PPS at periodicity $p$ of a real number signal.}
         \label{algo:max}
         \end{algorithm}
 %The periodic transform has the following properties. How about the Parseval's theorem?:
  %\begin{itemize}
 %          \item Commutativity
 %          \item Additivity
  %         \item Scalar Multiplication
 % \end{itemize}
 
\subsection{Algorithm for computing periodic power spectrum of a DNA sequence}
% File: getPSFromProfile.m
% Test file: DNAProfile_DFT5Profile_06302014
 From the definition of DFT power spectrum and the theorem 1, we then get the power spectrum of a DNA sequence after converting the sequence to 4D binary indicator sequences. For example, the power spectrum to describe 3-base periodicity property of protein coding regions in a DNA sequence can be obtained by its four congruence derivative vectors of the DNA sequence, which does not need to perform Fourier transform \citep{yin2007prediction}. Here, using PPS for 1D real signal, we propose the following algorithm to compute the power spectrum at any periodicities using the congruent derivative vectors  $f_\alpha  ,\alpha  \in \{ A,T,C,G\}$ at periodic positions. We propose following algorithm (Algorithm $2$) to compute the power spectrum at any periodicities in a DNA sequence. The inputs to the PPS algorithm are DNA sequence and the spectrum transform matrix $S_p$ to compute power spectrum at periodicity $p$ from the congruent derivative vectors. 
 
 \begin{algorithm}
 \DontPrintSemicolon % Some LaTeX compilers require to use \dontprintsemicolon instead
 \KwIn{DNA sequence of length $N$, periodicity $p$}
 \KwOut{PPS at periodicity $p$}
  \textbf{Step:}
  \begin{enumerate}
   \item Convert the DNA sequence into four binary indicator sequences $u_\alpha  ,\alpha  \in \{ A,T,C,G\}$.  
   
   \item Compute the congruent derivative vectors $f_{\alpha }$ of the sequences $u_\alpha  ,\alpha  \in \{ A,T,C,G\}$, at periodicity $p$.
   
    \item Compute the spectrum transform matrix $S_p$ of size $p$ ( refer to algorithm 1, steps 1-4).
   
   \item Compute the power spectrum  $PPS(p)$ by the four congruent derivative vectors  $f_{\alpha }$ of periodicity $p$:
   
   %\begin{equation}
  % MathType!MTEF!2!1!+-
  % feaafiart1ev1aaatCvAUfeBSjuyZL2yd9gzLbvyNv2CaerbuLwBLn
  % hiov2DGi1BTfMBaeXatLxBI9gBaerbd9wDYLwzYbItLDharqqtubsr
  % 4rNCHbWexLMBbXgBd9gzLbvyNv2CaeHbl7mZLdGeaGqiVCI8FfYJH8
  % YrFfeuY-Hhbbf9v8qqaqFr0xc9pk0xbba9q8WqFfeaY-biLkVcLq-J
  % Hqpepeea0-as0Fb9pgeaYRXxe9vr0-vr0-vqpWqaaeaabiGaciaaca
  % qabeaadaqaaqaafaGcbaGaamiuaiaadcfacaWGtbGaaiikaiaadcha
  % caGGPaGaeyypa0ZaaabuaeaacaWGMbWaaSbaaSqaaiabeg7aHbqaba
  % GccqGHxiIkcaWGtbWaaSbaaSqaaiaadchaaeqaaOGaey4fIOIaamOz
  % amaaDaaaleaacqaHXoqyaeaacaWGubaaaaqaaiabeg7aHjabgIGiol
  % aacUhacaWGbbGaaiilaiaadsfacaGGSaGaam4qaiaacYcacaWGhbGa
  % aiyFaaqab0GaeyyeIuoaaaa!5B92!
  \[
  PPS(p) = \sum\limits_{\alpha  \in \{ A,T,C,G\} } {f_\alpha   * S_p  * f_\alpha ^T } 
  \]
  % \end{equation}
   where ${f_{\alpha}^T }$ is the transpose of congruent derivative vector $f_{\alpha}$.
  \end{enumerate}
  
 \caption{Algorithm for computing the PPS at periodicity $p$ of a DNA sequence.}
 \label{algo:max}
 \end{algorithm}
 
 After we obtain the spectrum transform matrix $S{p}$ and the congruent derivative vectors from the algorithm 2 steps $1-3$, let $ f_{\alpha _0 } ,f_{\alpha _1 ,}  \cdots ,f_{\alpha _{p - 1} } $, $\alpha  \in \{ A,T,C,G\}$ be the elements of the congruent derivative vector, $f_{\alpha}$, of periodicity $p$ in a DNA sequence, which are the occurring frequency of nucleotide $\alpha$, we present Algorithm 2  to compute the $PPS(p)$ at periodicity $p$ of the DNA sequence. The $PPS(p)$ can also be computed as follows:
\begin{equation}
% MathType!MTEF!2!1!+-
% feaafiart1ev1aaatCvAUfeBSjuyZL2yd9gzLbvyNv2CaerbuLwBLn
% hiov2DGi1BTfMBaeXatLxBI9gBaerbd9wDYLwzYbItLDharqqtubsr
% 4rNCHbWexLMBbXgBd9gzLbvyNv2CaeHbl7mZLdGeaGqiVCI8FfYJH8
% YrFfeuY-Hhbbf9v8qqaqFr0xc9pk0xbba9q8WqFfeaY-biLkVcLq-J
% Hqpepeea0-as0Fb9pgeaYRXxe9vr0-vr0-vqpWqaaeaabiGaciaaca
% qabeaadaqaaqaafaGcbaGaamiuaiaadofacaWGqbGaaiikaiaadcha
% caGGPaGaeyypa0ZaaabuaeaadaqadaqaamaaqahabaGaamOzamaaDa
% aaleaacaWGHbWaaSbaaWqaaiaadghaaeqaaaWcbaGaaGOmaaaaaeaa
% caWGXbGaeyypa0JaaGimaaqaaiaadchacqGHsislcaaIXaaaniabgg
% HiLdGccqGHRaWkdaaeWbqaaiaadofadaWgaaWcbaGaamiCaaqabaGc
% caGGOaGaam4AaiaacYcacaWGQbGaaiykaiaadAgadaWgaaWcbaGaeq
% ySde2aaSbaaWqaaiaadQgaaeqaaaWcbeaakiaadAgadaWgaaWcbaGa
% eqySde2aaSbaaWqaaiaadUgaaeqaaaWcbeaaaeaacaWGRbGaeyypa0
% JaaGimaiaacYcacaWGQbGaeyypa0JaaGimaiaacYcacaWGRbGaeyOp
% a4JaamOAaaqaaiaadchacqGHsislcaaIXaaaniabggHiLdaakiaawI
% cacaGLPaaaaSqaaiabeg7aHjabgIGiopaacmaabaGaamyqaiaacYca
% caWGubGaaiilaiaadoeacaGGSaGaam4raaGaay5Eaiaaw2haaaqab0
% GaeyyeIuoaaaa!7BB8!
%\[
PPS(p) = \sum\limits_{\alpha  \in \left\{ {A,T,C,G} \right\}} {\left( {\sum\limits_{q = 0}^{p - 1} {f_{a_q }^2 }  + \sum\limits_{k = 0,j = 0,k > j}^{p - 1} {S_p (k,j)f_{\alpha _j } f_{\alpha _k } } } \right)} 
%\]
\end{equation}

 Because short periodicities often receive much attention in genome study, we provide the formulas for calculating the PPS spectrum of short periodicities, we construct the spectrum transform matrices for special short periodicities, $S_{2}$, $S_{3}$, $S_{4}$, $S_{5}$, $S_{6}$, as in Table 1 in supplementary materials. We then have following formulas for the PPS spectrum at a few short periodicities. 
 \begin{equation}
   % MathType!MTEF!2!1!+-
  % feaafiart1ev1aaatCvAUfeBSjuyZL2yd9gzLbvyNv2CaerbuLwBLn
  % hiov2DGi1BTfMBaeXatLxBI9gBaerbd9wDYLwzYbItLDharqqtubsr
  % 4rNCHbWexLMBbXgBd9gzLbvyNv2CaeHbl7mZLdGeaGqiVCI8FfYJH8
  % YrFfeuY-Hhbbf9v8qqaqFr0xc9pk0xbba9q8WqFfeaY-biLkVcLq-J
  % Hqpepeea0-as0Fb9pgeaYRXxe9vr0-vr0-vqpWqaaeaabiGaciaaca
  % qabeaadaqaaqaafaGcbaGaamiuaiaadcfacaWGtbGaaiikaiaaikda
  % caGGPaGaeyypa0ZaaabuaeaadaqadaqaaiaadAgadaqhaaWcbaGaeq
  % ySde2aaSbaaWqaamaaBaaabaGaaGimaaqabaaabeaaaSqaaiaaikda
  % aaGccqGHRaWkcaWGMbWaa0baaSqaaiabeg7aHnaaBaaameaadaWgaa
  % qaaiaaigdaaeqaaaqabaaaleaacaaIYaaaaOGaeyOeI0IaaGOmaiaa
  % dAgadaWgaaWcbaGaeqySde2aaSbaaWqaamaaBaaabaGaaGimaaqaba
  % aabeaaaSqabaGccaWGMbWaaSbaaSqaaiabeg7aHnaaBaaameaadaWg
  % aaqaaiaaigdaaeqaaaqabaaaleqaaaGccaGLOaGaayzkaaaaleaacq
  % aHXoqycqGHiiIZdaGadaqaaiaadgeacaGGSaGaamivaiaacYcacaWG
  % dbGaaiilaiaadEeaaiaawUhacaGL9baaaeqaniabggHiLdaaaa!6640!
  %\[
  PPS(2) = \sum\limits_{\alpha  \in \left\{ {A,T,C,G} \right\}} {\left( {f_{\alpha _{_0 } }^2  + f_{\alpha _{_1 } }^2  - 2f_{\alpha _{_0 } } f_{\alpha _{_1 } } } \right)} 
  %\]
 \end{equation}
  \begin{equation}
% MathType!MTEF!2!1!+-
% feaafiart1ev1aaatCvAUfeBSjuyZL2yd9gzLbvyNv2CaerbuLwBLn
% hiov2DGi1BTfMBaeXatLxBI9gBaerbd9wDYLwzYbItLDharqqtubsr
% 4rNCHbWexLMBbXgBd9gzLbvyNv2CaeHbl7mZLdGeaGqiVCI8FfYJH8
% YrFfeuY-Hhbbf9v8qqaqFr0xc9pk0xbba9q8WqFfeaY-biLkVcLq-J
% Hqpepeea0-as0Fb9pgeaYRXxe9vr0-vr0-vqpWqaaeaabiGaciaaca
% qabeaadaqaaqaafaGcbaGaamiuaiaadofacaWGqbGaaiikaiaaioda
% caGGPaGaeyypa0ZaaabuaeaadaqadaqaamaaqahabaGaamOzamaaDa
% aaleaacaWGHbWaaSbaaWqaaiaadghaaeqaaaWcbaGaaGOmaaaaaeaa
% caWGXbGaeyypa0JaaGimaaqaaiaaikdaa0GaeyyeIuoakiabgkHiTi
% aadAgadaWgaaWcbaGaeqySde2aaSbaaWqaaiaaicdaaeqaaaWcbeaa
% kiaadAgadaWgaaWcbaGaeqySde2aaSbaaWqaaiaaigdaaeqaaaWcbe
% aakiabgkHiTiaadAgadaWgaaWcbaGaeqySde2aaSbaaWqaaiaaicda
% aeqaaaWcbeaakiaadAgadaWgaaWcbaGaeqySde2aaSbaaWqaaiaaik
% daaeqaaaWcbeaakiabgkHiTiaadAgadaWgaaWcbaGaeqySde2aaSba
% aWqaaiaaigdaaeqaaaWcbeaakiaadAgadaWgaaWcbaGaeqySde2aaS
% baaWqaaiaaikdaaeqaaaWcbeaaaOGaayjkaiaawMcaaaWcbaGaeqyS
% deMaeyicI48aaiWaaeaacaWGbbGaaiilaiaadsfacaGGSaGaam4qai
% aacYcacaWGhbaacaGL7bGaayzFaaaabeqdcqGHris5aaaa!7582!
%\[
PPS(3) = \sum\limits_{\alpha  \in \left\{ {A,T,C,G} \right\}} {\left( {\sum\limits_{q = 0}^2 {f_{a_q }^2 }  - f_{\alpha _0 } f_{\alpha _1 }  - f_{\alpha _0 } f_{\alpha _2 }  - f_{\alpha _1 } f_{\alpha _2 } } \right)} 
%\]
\end{equation}  
  \begin{equation}  
% MathType!MTEF!2!1!+-
% feaafiart1ev1aaatCvAUfeBSjuyZL2yd9gzLbvyNv2CaerbuLwBLn
% hiov2DGi1BTfMBaeXatLxBI9gBaerbd9wDYLwzYbItLDharqqtubsr
% 4rNCHbWexLMBbXgBd9gzLbvyNv2CaeHbl7mZLdGeaGqiVCI8FfYJH8
% YrFfeuY-Hhbbf9v8qqaqFr0xc9pk0xbba9q8WqFfeaY-biLkVcLq-J
% Hqpepeea0-as0Fb9pgeaYRXxe9vr0-vr0-vqpWqaaeaabiGaciaaca
% qabeaadaqaaqaafaGcbaGaamiuaiaadofacaWGqbGaaiikaiaaisda
% caGGPaGaeyypa0ZaaabuaeaadaqadaqaamaaqahabaGaamOzamaaDa
% aaleaacqaHXoqydaWgaaadbaGaamyCaaqabaaaleaacaaIYaaaaaqa
% aiaadghacqGH9aqpcaaIWaaabaGaaG4maaqdcqGHris5aOGaeyOeI0
% IaaGOmaiaadAgadaWgaaWcbaGaeqySde2aaSbaaWqaaiaaicdaaeqa
% aaWcbeaakiaadAgadaWgaaWcbaGaeqySde2aaSbaaWqaaiaaikdaae
% qaaaWcbeaakiabgkHiTiaaikdacaWGMbWaaSbaaSqaaiabeg7aHnaa
% BaaameaacaaIXaaabeaaaSqabaGccaWGMbWaaSbaaSqaaiabeg7aHn
% aaBaaameaacaaIZaaabeaaaSqabaaakiaawIcacaGLPaaaaSqaaiab
% eg7aHjabgIGiopaacmaabaGaamyqaiaacYcacaWGubGaaiilaiaado
% eacaGGSaGaam4raaGaay5Eaiaaw2haaaqab0GaeyyeIuoaaaa!6F64!
%\[
PPS(4) = \sum\limits_{\alpha  \in \left\{ {A,T,C,G} \right\}} {\left( {\sum\limits_{q = 0}^3 {f_{\alpha _q }^2 }  - 2f_{\alpha _0 } f_{\alpha _2 }  - 2f_{\alpha _1 } f_{\alpha _3 } } \right)} 
%\]
 \end{equation}

The signal-to-noise ratio (SNR) of a DNA sequence at a periodicity $p$, is defined as its PPS power spectrum at periodicity $p$ divided by the average DFT power spectrum. From our previous study, it was found that the average DFT power spectrum corresponds to the sequence length $N$ \citep{yin2007prediction}. Thus the SNR of PPS spectrum a DNA sequence at periodicity $p$ is defined as the PPS power spectrum at periodicity $p$ divided by the length of the DNA sequence.
  \begin{equation}
 % MathType!MTEF!2!1!+-
 % feaafiart1ev1aaatCvAUfeBSjuyZL2yd9gzLbvyNv2CaerbuLwBLn
 % hiov2DGi1BTfMBaeXatLxBI9gBaerbd9wDYLwzYbItLDharqqtubsr
 % 4rNCHbWexLMBbXgBd9gzLbvyNv2CaeHbl7mZLdGeaGqiVCI8FfYJH8
 % YrFfeuY-Hhbbf9v8qqaqFr0xc9pk0xbba9q8WqFfeaY-biLkVcLq-J
 % Hqpepeea0-as0Fb9pgeaYRXxe9vr0-vr0-vqpWqaaeaabiGaciaaca
 % qabeaadaqaaqaafaGcbaGaam4uaiaad6eacaWGsbGaaiikaiaadcha
 % caGGPaGaeyypa0ZaaSaaaeaacaWGqbGaamiuaiaadofacaGGOaGaam
 % iCaiaacMcaaeaacaWGobaaaaaa!4A96!
 %\[
 SNR(p) = \frac{{PPS(p)}}
 {N}
 %\]
  \end{equation}
  In this paper, we chose the threshold of SNR as 1 to differ a true periodicity and background noise.
 
\section{Results and Discussions}
\subsection{PPS power spectrum analysis of a periodic signal}
%Yin0216
To illustrate the effectiveness of the PPS algorithm in the identification of hidden periodicities in signals, we applied the PPS algorithm to the following periodic signal that consists of sine and cosine signal with periodicities 20 and 50, and is corrupted by white Gaussian noise:
% MathType!MTEF!2!1!+-
% feaafiart1ev1aaatCvAUfeBSjuyZL2yd9gzLbvyNv2CaerbuLwBLn
% hiov2DGi1BTfMBaeXatLxBI9gBaerbd9wDYLwzYbItLDharqqtubsr
% 4rNCHbWexLMBbXgBd9gzLbvyNv2CaeHbl7mZLdGeaGqiVCI8FfYJH8
% YrFfeuY-Hhbbf9v8qqaqFr0xc9pk0xbba9q8WqFfeaY-biLkVcLq-J
% Hqpepeea0-as0Fb9pgeaYRXxe9vr0-vr0-vqpWqaaeaabiGaciaaca
% qabeaadaqaaqaafaGceaqabeaacaWG4bGaaiikaiaad6gacaGGPaGa
% eyypa0Jaci4CaiaacMgacaGGUbGaaiikaiaaikdacqaHapaCdaWcaa
% qaaiaad6gaaeaacaaIYaGaaGimaaaacqGHRaWkdaWcaaqaaiabec8a
% WbqaaiaaisdaaaGaaiykaiabgUcaRiGacogacaGGVbGaai4CaiaacI
% cacaaIYaGaeqiWda3aaSaaaeaacaWGUbaabaGaaGynaiaaicdaaaGa
% ey4kaSYaaSaaaeaacqaHapaCaeaacaaI0aaaaiaacMcacqGHRaWkca
% WGUbGaam4BaiaadMgacaWGZbGaamyzaaqaaiaad6gacqGH9aqpcaaI
% XaGaaiOoaiaaiodacaaIWaGaaGimaaaaaa!689F!
\[
\begin{gathered}
  x(n) = \sin (2\pi \frac{n}
{{20}} + \frac{\pi }
{4}) + \cos (2\pi \frac{n}
{{50}} + \frac{\pi }
{4}) + noise \hfill \\
  n = 1:300 \hfill \\ 
\end{gathered} 
\]
%Figure 1
Figure 1(a) is the plot of the original periodic signal and shows that two periodicities 20 and 50 in the signal are hidden by random noise. Figure 1(b) is the plot of Fourier power spectrum vs frequency and shows that the two periodicities can be clearly identified by Fourier transform. The Fourier power spectrum plotting is over frequency domain, thus the positions of the two periodicity peaks 20 and 50 are at frequency $N/p$, i.e., frequency 15 and 6, respectively. Figure 1(c) is the plot of PPS spectrum vs periodicity. Two pronounced peaks of PPS spectrum in Figure 1(c) at positions 20 and 50 represent the two periodicities hidden in the original signal. The strengths of the periodicity 20 and 50 from the Fourier power spectrum also equal to the strengths of the corresponding PPS spectrum.  signals from the  In comparison of the spectra of Fourier transform and PPS, we can see that PPS spectrum can be used to identify periodicities directly in frequency domain. In addition, from Figure 1(b) and (a), we can see the PPS spectrum has less noise background than the Fourier power spectrum.
 %File:DFT_Protein/getPSFromSignalFraction_Test_02102015.m
  \begin{figure}[tbp]
      \centering
      \subfloat[]{\includegraphics[width=3.25in]{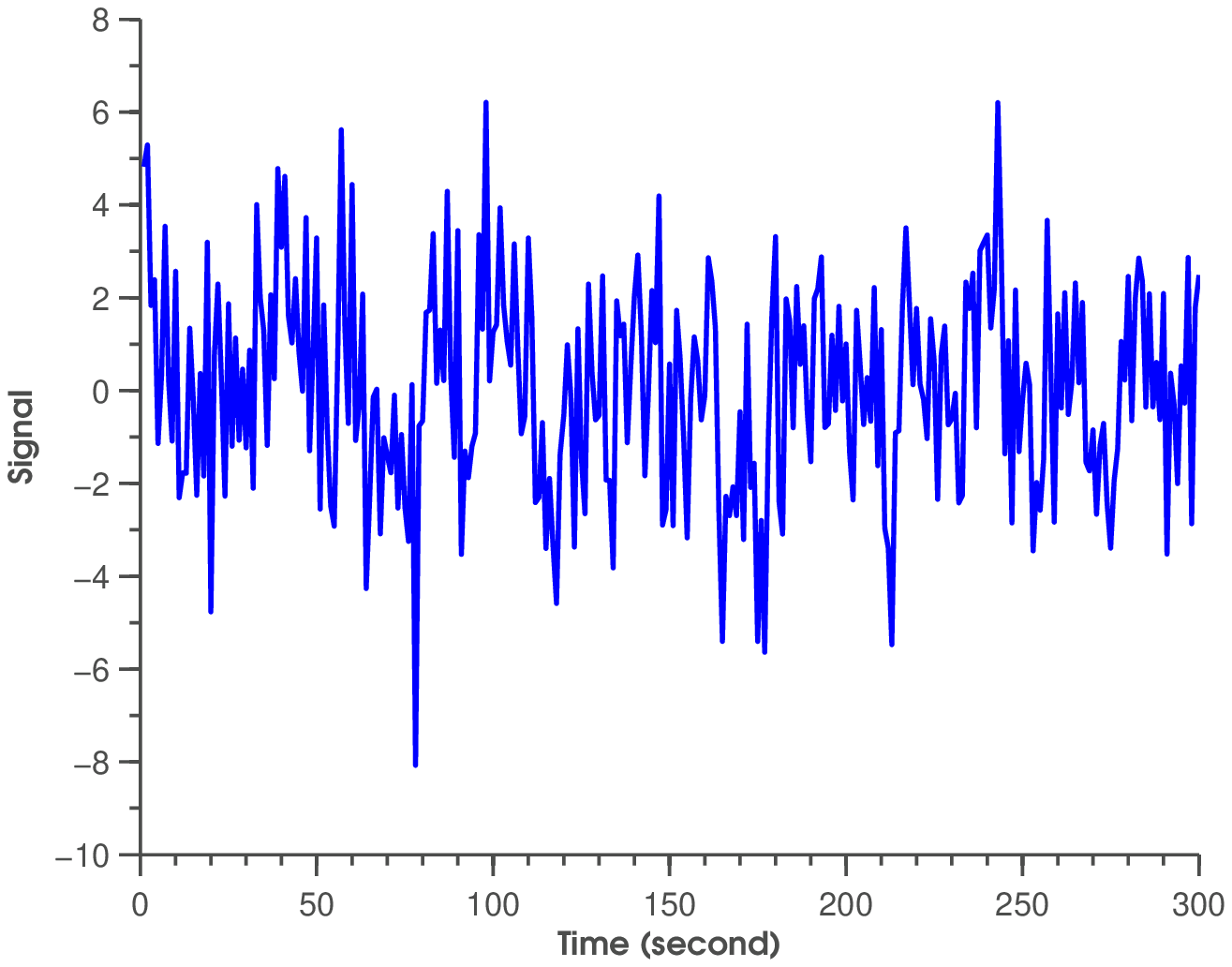}}\quad
      \subfloat[]{\includegraphics[width=3.25in]{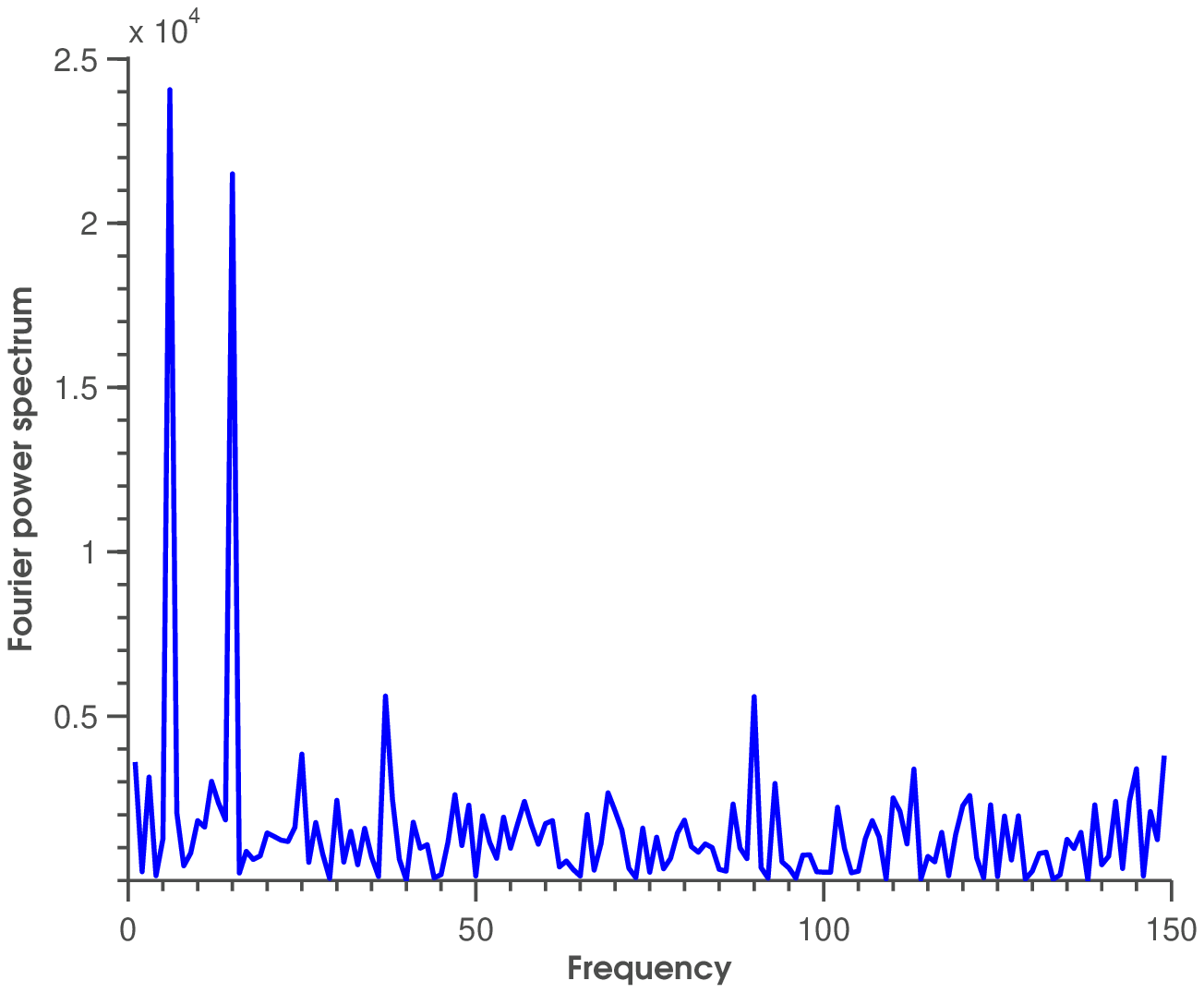}}\quad
      \subfloat[]{\includegraphics[width=3.25in]{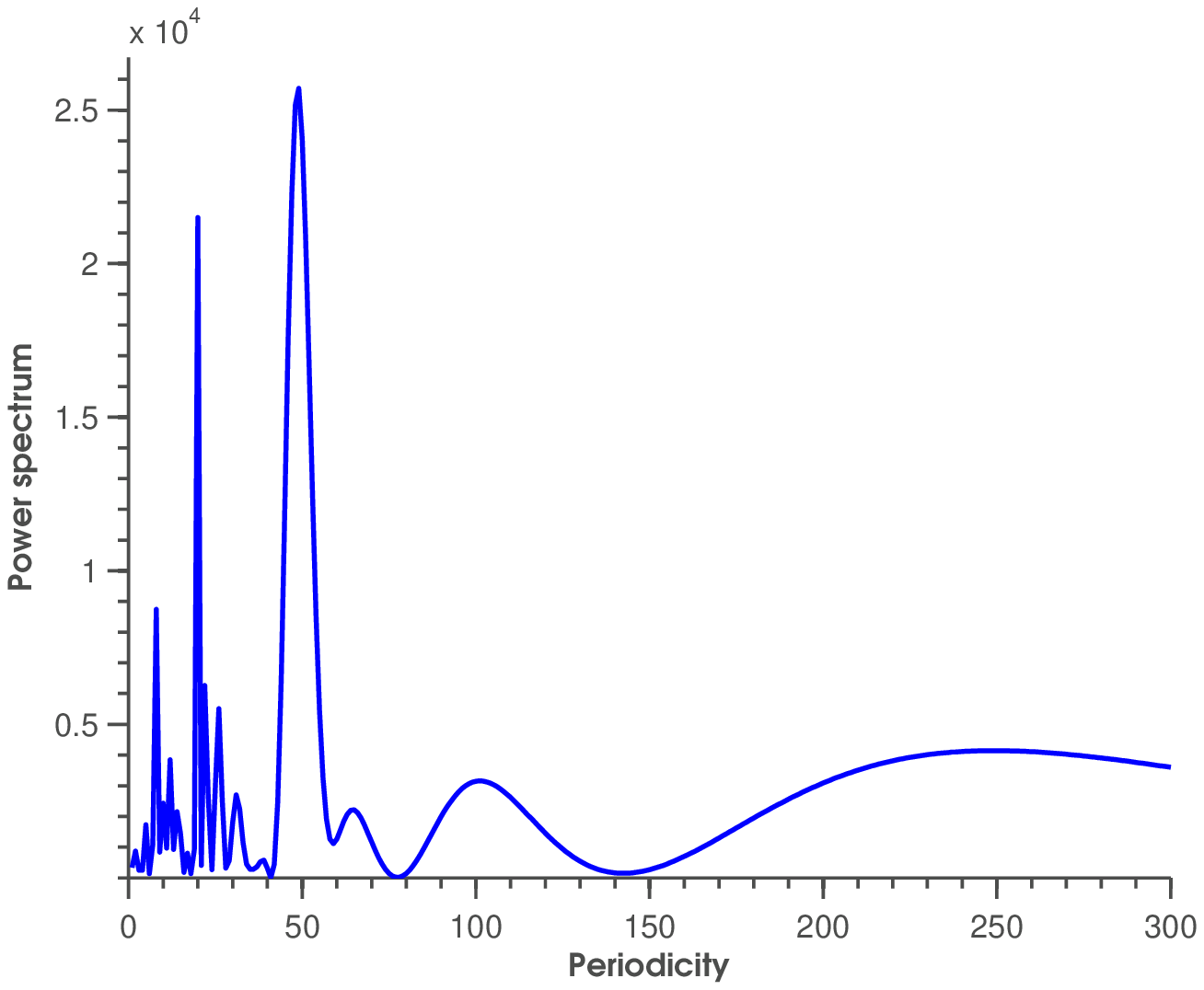}}\quad
      \caption{Power spectrum analysis of periodic sine and cosine signal. (a) Periodic sine and cosine signal collapsed with white noise, (b) DFT power spectrum of the signal, (c) PPS power spectrum of the signal.}
      \label{fig:sub1}
    \end{figure}
   
  \subsection{PPS power spectrum analysis of DNA sequences} 
 %YIN0216
 We also studied on the impact of spectral leakages on Fourier power spectrum and PPS spectrum. For periodic sinusoidal signals, $A= \sin (2\pi f_i nT)$, where $N$ is the number of samples and $T$ is the distance between two samples, and  $n =0, 1, 2,... N-1$. The Fourier power spectrum of the signal has a peak proportional to amplitude at the frequency $f_i$. If the frequency $f_i$ of the signal is not multiple of $F,F = 1/NT$, a distorted DFT is obtained. This phenomenon is called spectral leakage \citep{costa2010information,lyon2009discrete}. Because spectral leakage makes the recognition of the correct frequencies of the signal difficult, it shall be avoided in digital signal processing. To avoid spectral leakage at a specific periodicity, we can pad zeros to make the length of DNA sequence as a multiple of the periodicity. For example, to compute accurate power spectrum at periodicity 3, the length of the DNA sequence after padding zeros shall be a multiple of 3. 
     
 %Figure 2 
 The test DNA sequence in the spectral leakage analysis is an artificial sequence, denoted as N130P5. The DNA sequence contains 6 copies imperfect 5-base repeat ATCGA. A deletion mutant of the DNA sequence is constructed by deleting two nucleotides AA at the 3' end, denoted as N130P5-D2. The sequences N130P5 and N130P5-D2 are provided in the supplementary materials. The Fourier power spectrum and PPS spectrum are shown in Figure 2 and Table 3. The Fourier power spectrum and PPS spectrum for the periodicity 5 are the same, the peak value is 361.9837 (Figure 2(a)(c)). The results verify that the  PPS spectrum at periodicity 5 is the same as that from the DFT method on the sequence when its length is multiple of periodicity 5. Figure 2(a) is the comparison of Fourier power spectrum of  N130P5 and N130P5-D2 with zero padding. The result in Figure 2(a) shows that zero padding for the deletion mutant can have similar spectrum as the original DNA sequence. However, if there is no zero padding in the  N130P5-D2 sequence as shown in Figure 2(b), the Fourier power spectrum at periodicity 5 becomes 212.0118 and is much different from the corresponding values from padding zeros (335.803) and PPS spectrum of original sequence (335.8034). The reason for these difference is that the length of the deletion mutant is not multiple of 5 and the periodicity 5 is hidden in Fourier spectrum because of spectral leakage. In addition, PPS spectrum can be considered as zero padding for each periodicity as shown from equation (7). These results indicate that Fourier spectrum may not identify some latent periodicities due to spectrum leakage, but the PPS spectrum can detect the hidden periodicity no matter whether the length of the sequence is multiple of the periodicity. This is the main advantage of the PPS spectrum. 
     %File:DNA_profile/DNAProfile_P5DEL_Zeros_01112015.m
     \begin{figure}[tbp]
       	\centering
       	\subfloat[]{\includegraphics[width=3.25in]{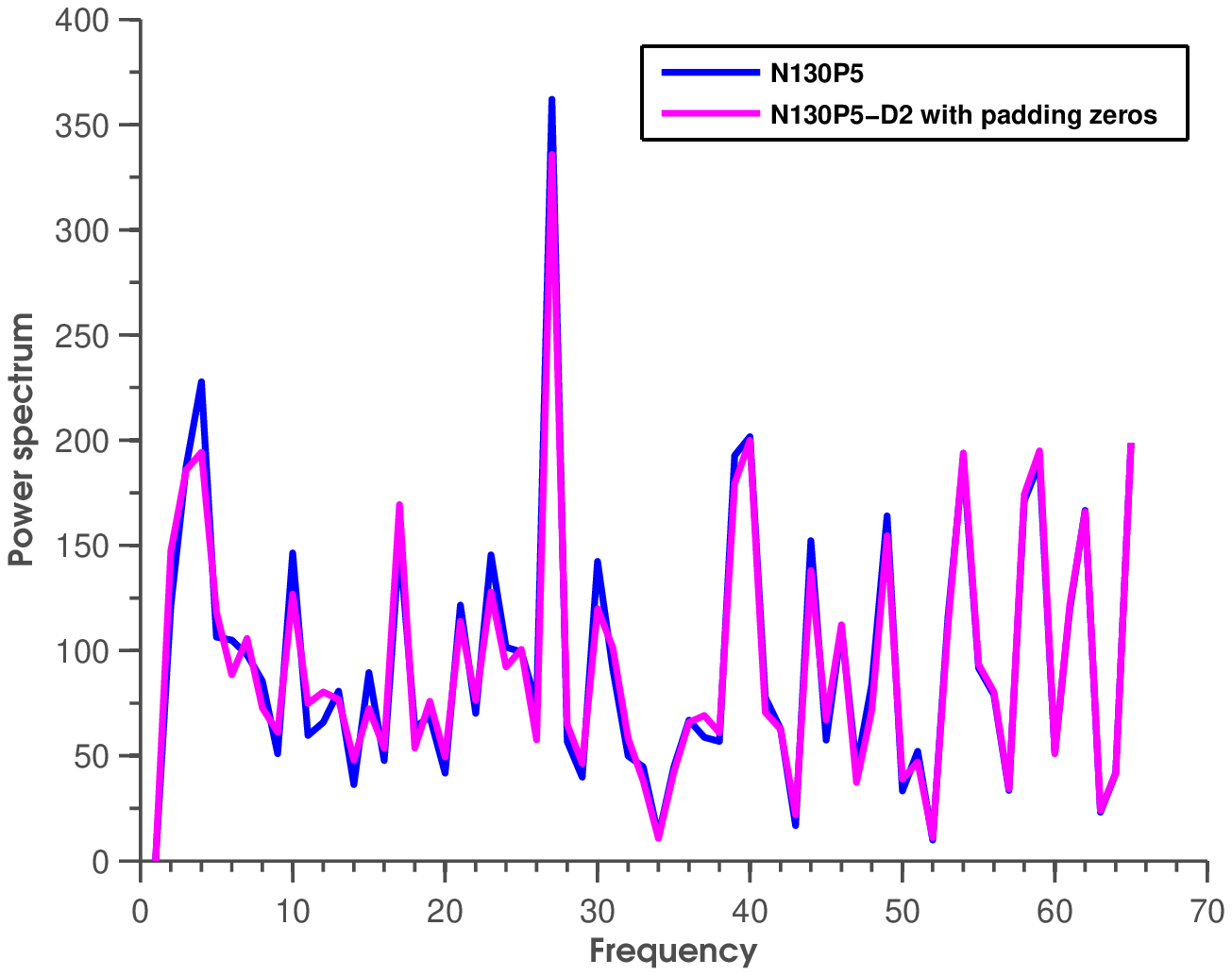}}\quad
       	\subfloat[]{\includegraphics[width=3.25in]{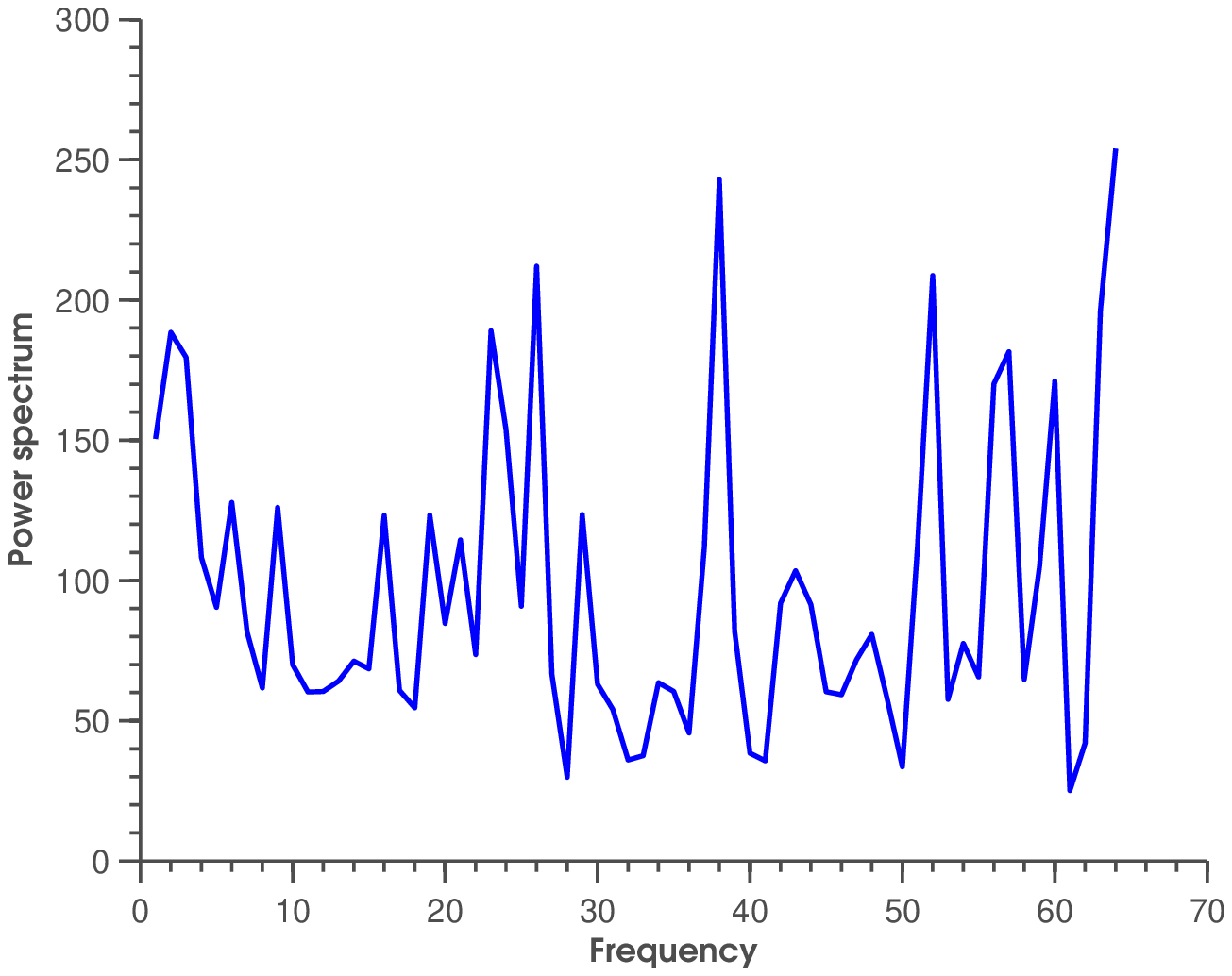}}\quad
       	\subfloat[]{\includegraphics[width=3.25in]{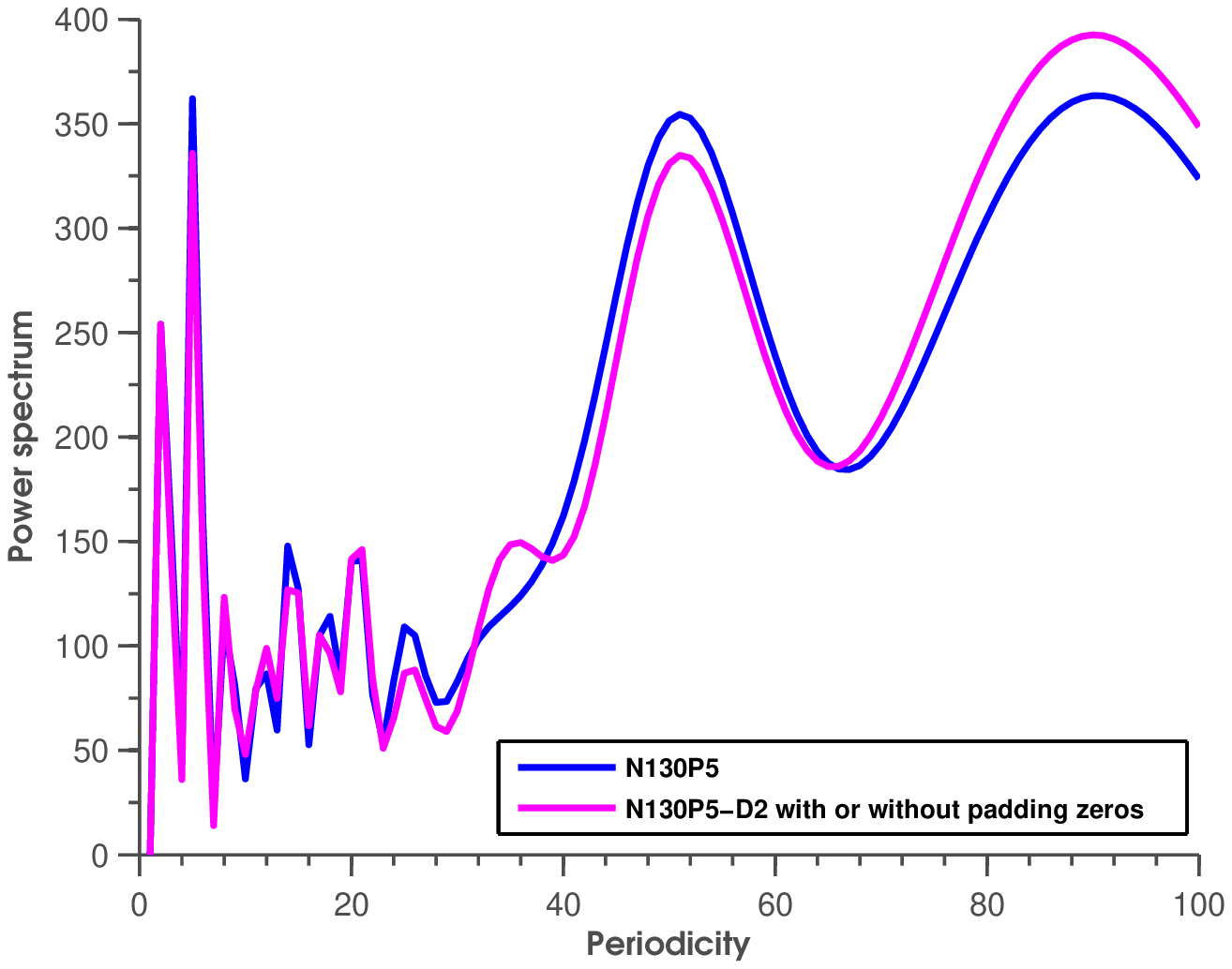}}\quad
        \caption{Power spectrum analysis of DNA sequences. (a) Fourier power spectra of DNA sequence N130P5 and N130P5-D2, (b) Fourier power spectrum of N130P5-D2, (c) PPS spectra of and N130P5 and N130P5-D2.}
       \end{figure}
       
        \begin{table}[ht]
        \caption{Comparison of Fourier power spectrum and PPS spectrum on simulated DNA sequences} % title of Table
        \centering % used for centering table
        \begin{tabular}{l l l} % centered columns (3 columns)
        \hline\hline %inserts double horizontal lines
        Sequence & DFT:$PS(\frac{N}
        {5})$  & PPS:$PS(5)$ \\ [0.5ex] % inserts table
        %heading
        \hline % inserts single horizontal line
        N130P5 & 361.9837 & 361.9837  \\ % inserting body of the table
        N130P5-D2 (with padding zeros) & 335.8034 & 335.8034  \\ % inserting body of the table
        N130P5-D2 (no padding zeros) & 212.0118 & 335.8034  \\ % inserting body of the table
        \hline\hline %inserts single line
        \end{tabular}
        \label{table:nonlin} % is used to refer this table in the text
        \end{table}
        
The PPS spectrum method was assessed on well-studied 3-periodicity of exon sequences. Figure 3(a) is Fourier power spectrum \textit{Homo sapiens} cytochrome oxidase subunit I (COI) gene (GenBank ID: EU834863, 617 bp). Figure 3(b) is the PPS spectrum and indicates a significant periodicities 3 and a weak periodicity 8 in the gene. We then examine these two periodicities using DNA walk and sliding window approaches as described in our previous study \citep{yin2007prediction}  Figure 3(c) is the PPS spectra of periodicities 3 and 8 using DNA walk. The strengths of the PPM spectrum for the two periodic signals increase when the length of the DNA walk increases. Figure 3(d) is the PPS spectra of periodicities 3 and 8 in different sliding windows. From the sliding window plots, we can identify the approximate regions in the gene for these two periodic signals. These results show that the weak periodicity 8 can only be identified in PPS spectrum, but not in Fourier power spectrum.
 \begin{figure}[tbp]
           \centering
           \subfloat[][]{\includegraphics[width=2.5in]{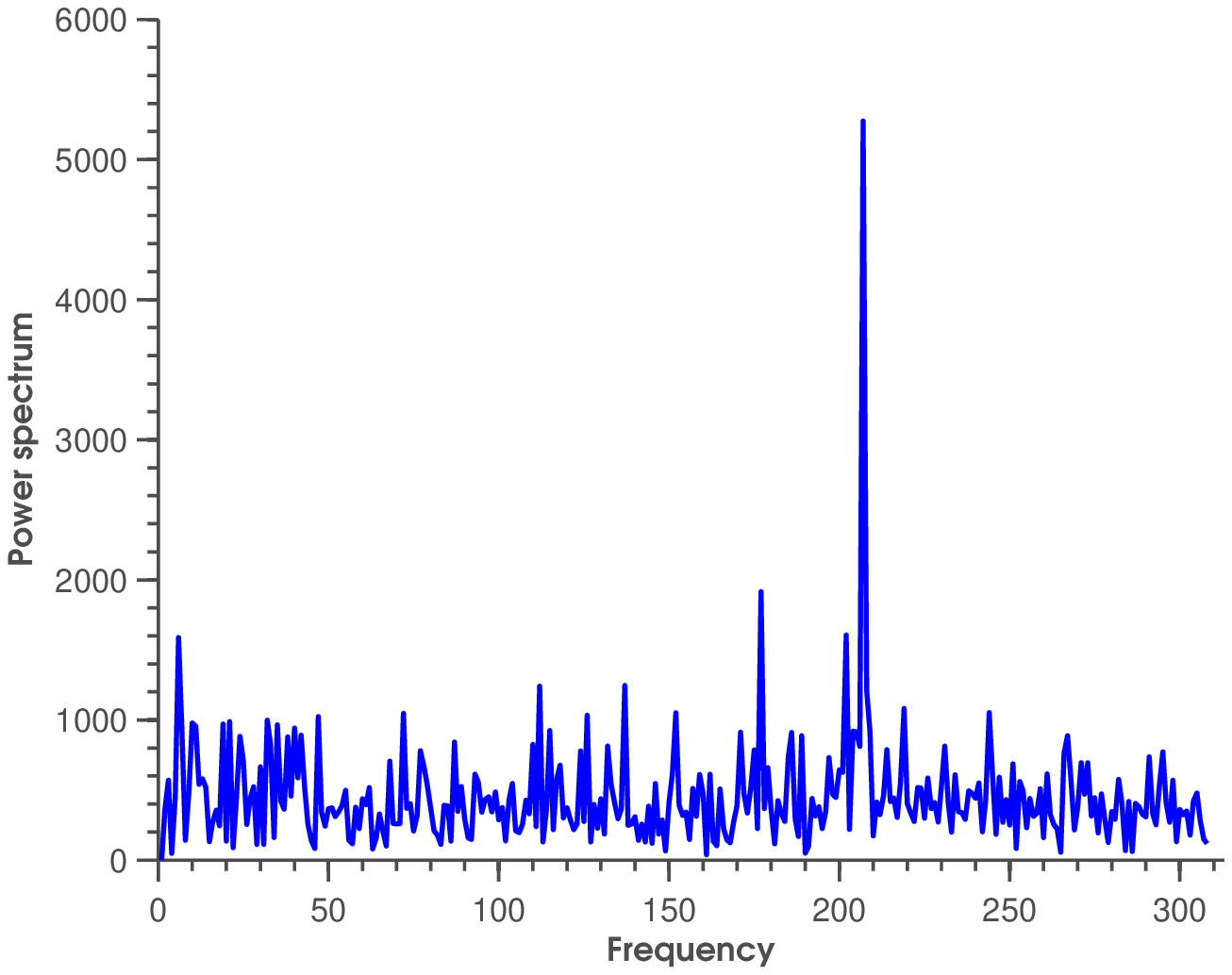}}\quad
           \subfloat[][]{\includegraphics[width=2.5in]{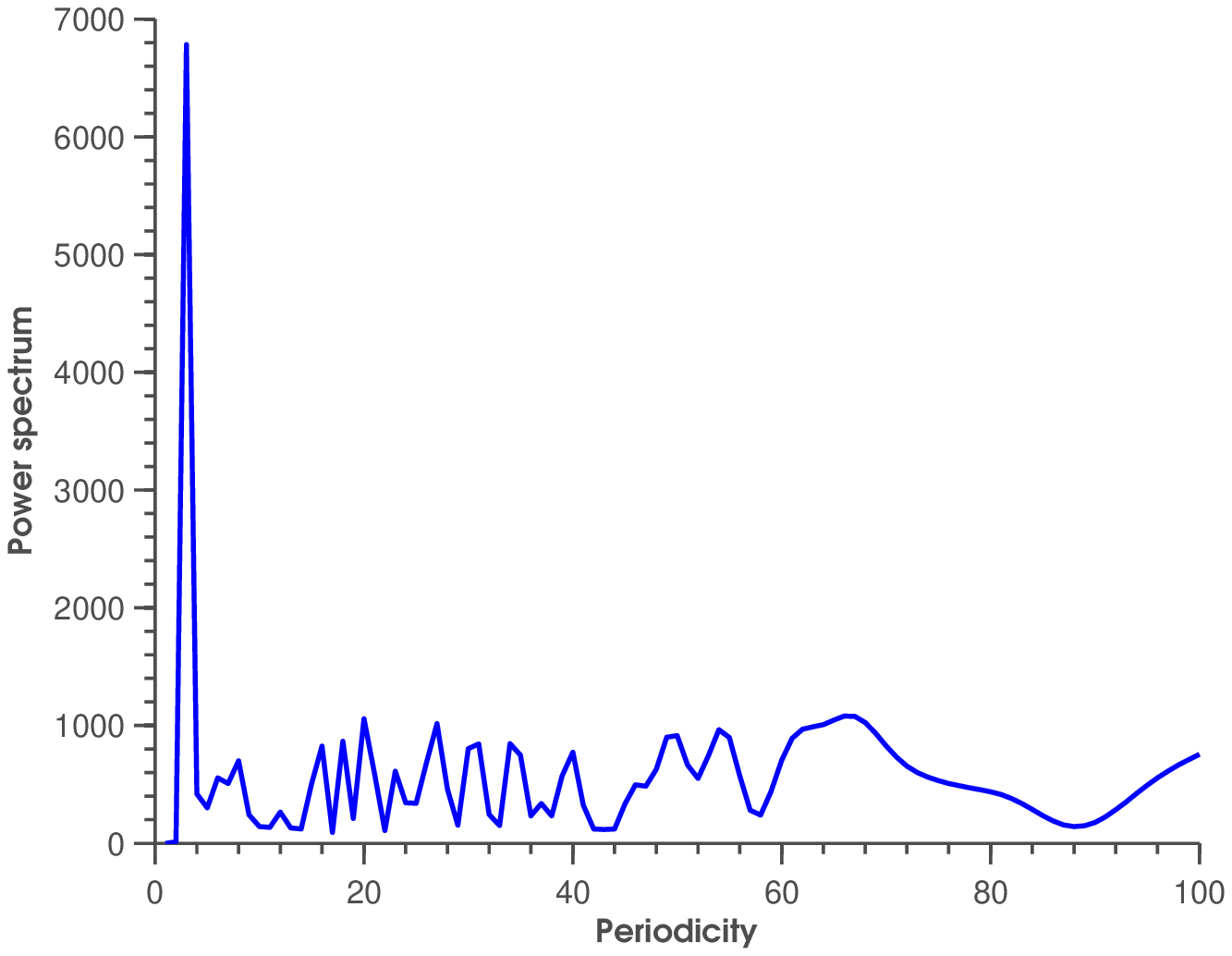}}\\
           \subfloat[][]{\includegraphics[width=2.5in]{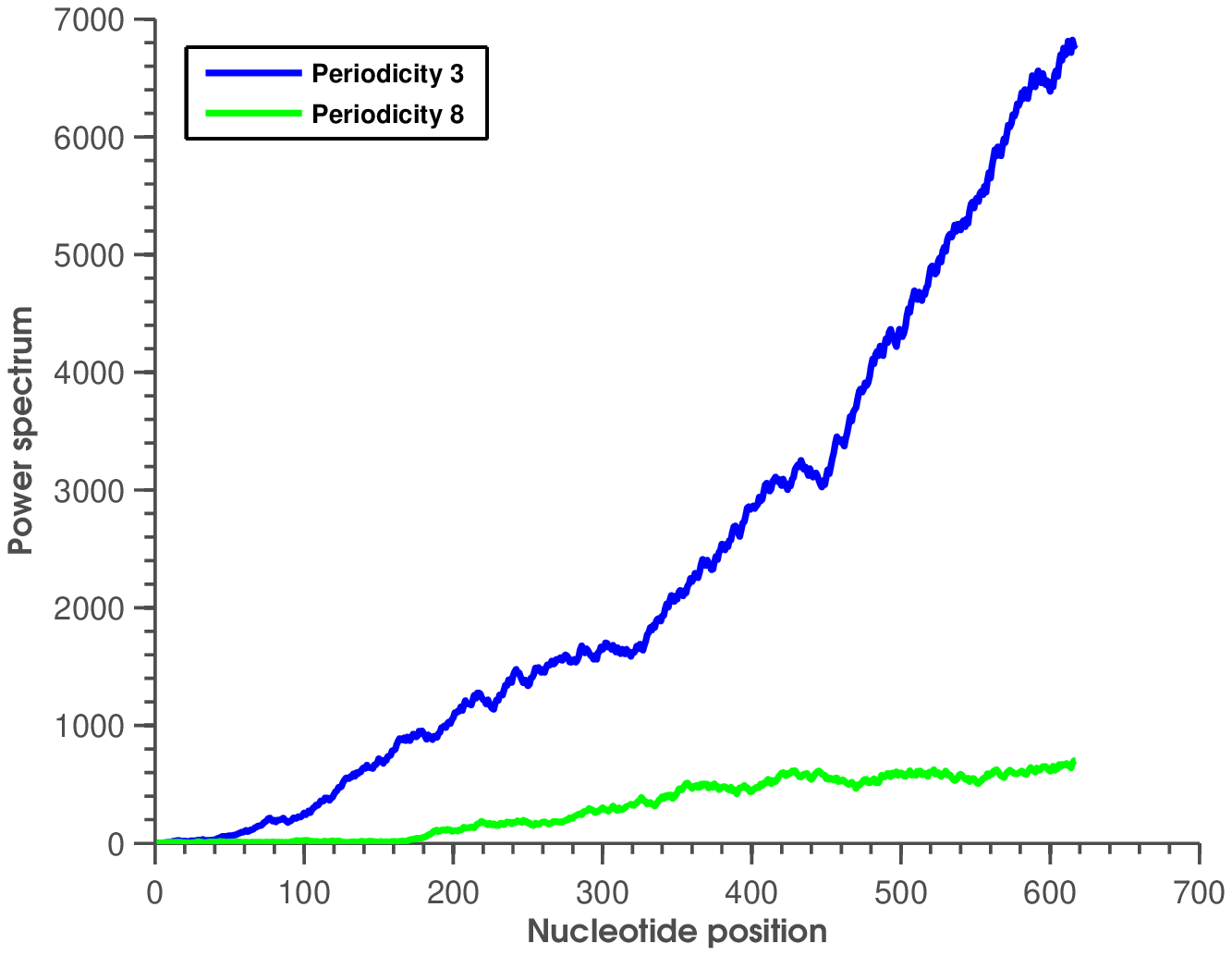}}\quad
           \subfloat[][]{\includegraphics[width=2.5in]{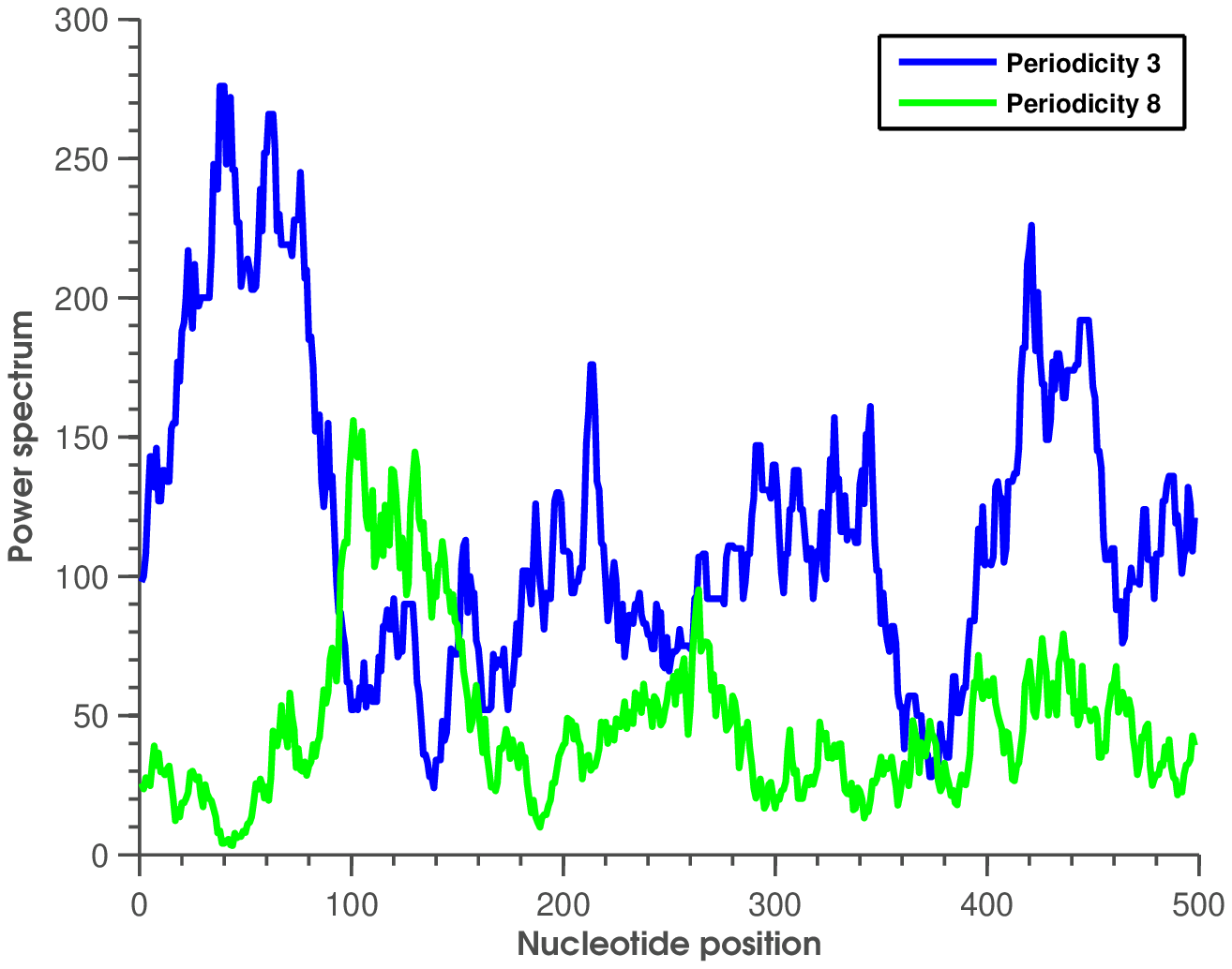}}
           \caption{Power spectra of \textit{Homo sapiens} cytochrome oxidase subunit 1 (COI) gene. (a) Fourier power spectrum, (b) PPS spectrum, (c) PPS spectra of DNA walk, (d) sliding window PPS spectrum, window size = 60 bp}
           \label{fig:sub1}
\end{figure}
        
We performed other tests on DNA sequences of different lengths, the computational results confirm that the PPS spectrum and Fourier power spectrum are the same for periodicity $p$ only when the length of the DNA sequences are the multiple of $p$. If the length of the DNA sequence is not a multiple of the periodicity $p$, the Fourier power spectrum is not equivalent to the PPS spectrum. This indicates that one periodic signal may be hidden by other periodicities in Fourier power spectrum analysis. This study shows that the Fourier power spectrum signal computation in DNA sequence is impacted by length of DNA sequences, but our new PPS spectrum can overcome the length problem as in Fourier spectrum analysis. 

\subsection{Periodicities in microsatellite sequences}
%Figure 3
We evaluated the effectiveness of the PPS method by comparing Fourier and PPS spectra of typical microsatellite DNA sequences.  The most studied groups of tandem repeats in genomes are microsatellites (patterns of 10 bp) and minisatellites (patterns of 100 bp) because of their use as genetic markers in forensics, parentage assessment, positional cloning and population and evolutionary genetics. Microsatellites are abundant and distributed all over the eukaryotic genomes with variable frequency. 
%http://www.ncbi.nlm.nih.gov/pmc/articles/PMC3171338/pdf/1687-4153-2007-43596.pdf
 %http://research.cs.wisc.edu/gensoft/TR/index.html (sequence information)
%Figure 4
The period detection method using PPS was assessed on Human microsatellite repeat KLK1 AC DNA (GenBank Accession: M65145, 1072 bp). Figure 4(a) shows the Fourier power spectrum of the DNA sequence. Although the Fourier power spectrum indicates there are two high peaks corresponding to approximate 11 and 12 periodicities, it is difficult to identify other exact periodicities from the Fourier power spectrum. Figure 4(b) is the PPS spectrum SNR at periodicities 1-100. It is clear that the DNA sequence contains a few periodicity peaks with large spectrum SNR values. The significant SNR values for short periodicities are: $P_7$:1.3873, $P_8$:1.2002, $P_{11}$:1.5389, $P_{12}$:2.4196. Our method can thus preciously identify 4 periodicities (repeats): 7, 8, 11, and 12. This sequence had been studied by other repeat finding methods \citep{sharma2004spectral,gupta2007novel}, but those methods can only detect 2 or 11 mer repeats and were unable to detect the four periodicities 7, 8, 11, and 12. We can locate the positions of the periodicities in the DNA sequence using sliding window PPS. Figure 4(c) is the sliding window PPS of periodicities 7, 11 and 12. The result shows the different locations for these periodic signals on the DNA sequence. The 11-mer repeats are located between 92 and 781 bp. When comparing the sliding window PPS SNR plot of 11 bp and 12 periodicities (Figure 4(b),(c)), we can see that the peak regions from 11 and 12 periodicity are not the same, but the region centered at 200 bp only shows 11 periodicity, but not 11 periodicity. This result suggest the 11 periodicity is not exact derived from 12 periodicity. The locations the periodicity 11 are in agreement of previous studies \citep{sharma2004spectral,gupta2007novel}. These results clearly show the advantage of our algorithm in identification of repeats. The PPS method can detect more latent periodicities and hidden periodic patterns and locate preciously the locations of these periodicities. 
 
 %File:DNA_Profile/DNAProfile_M65145_01082015.m
   \begin{figure}[tbp]
       \centering
       \subfloat[]{\includegraphics[width=3.25in]{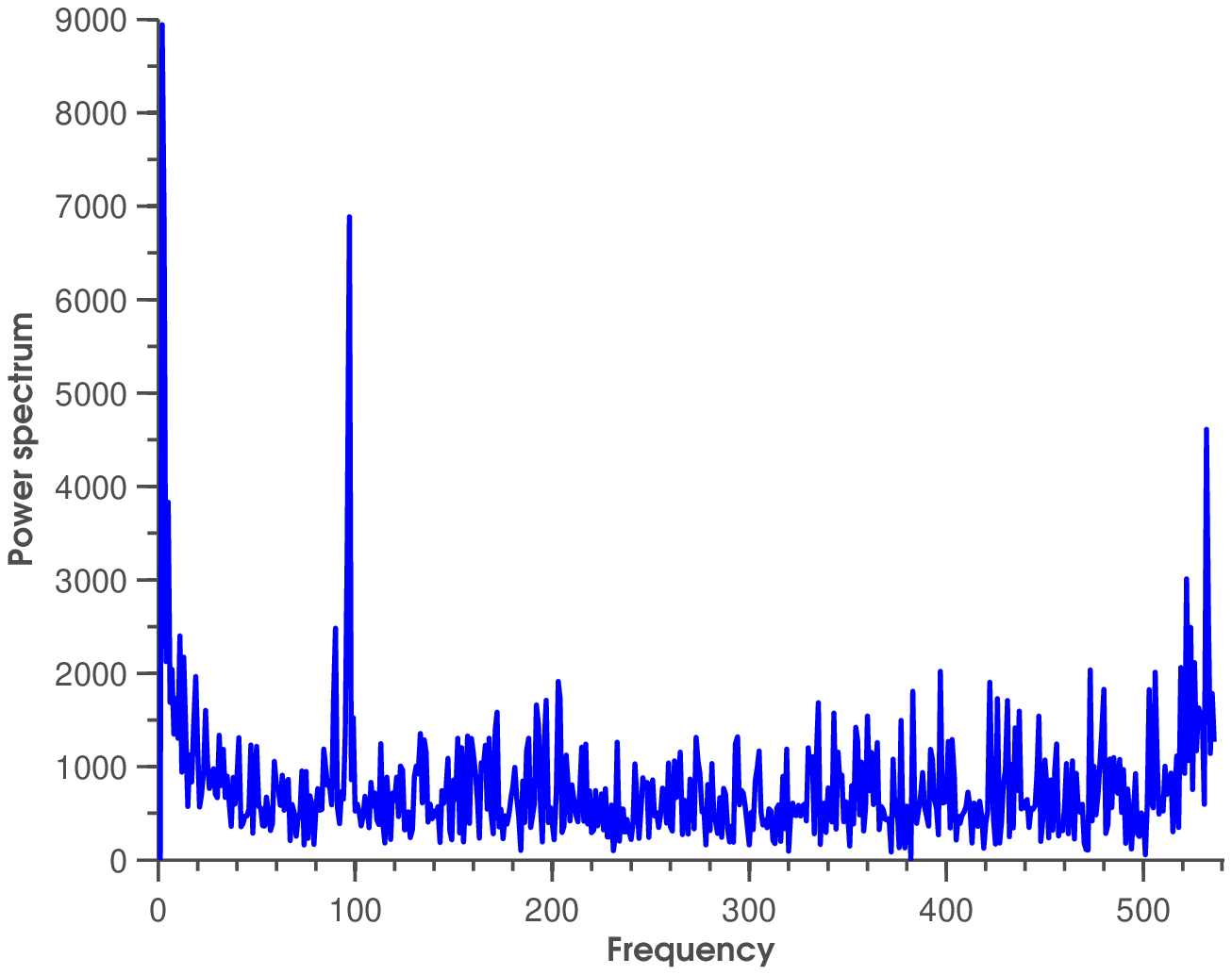}}\quad
       \subfloat[]{\includegraphics[width=3.25in]{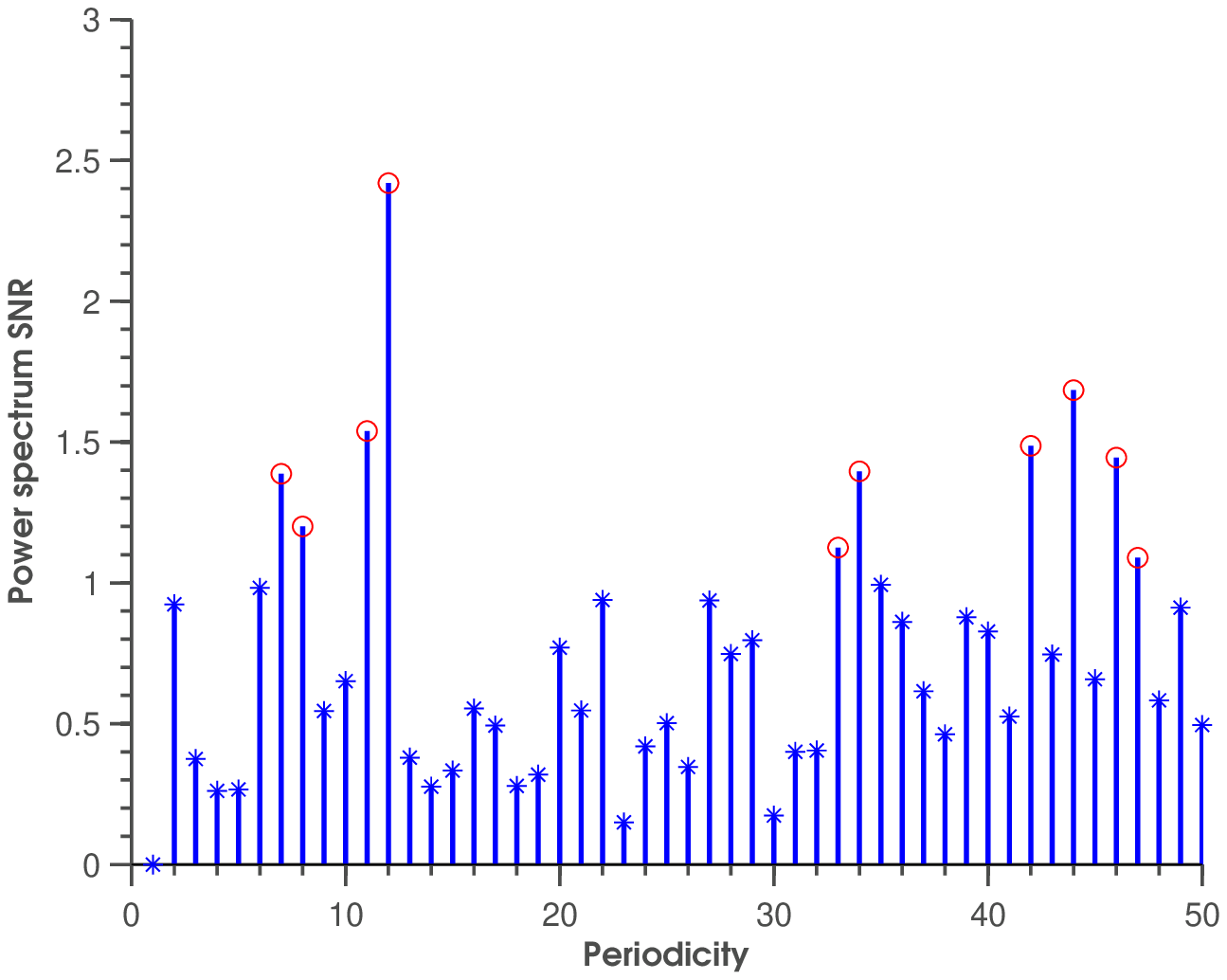}}\quad
       \subfloat[]{\includegraphics[width=3.25in]{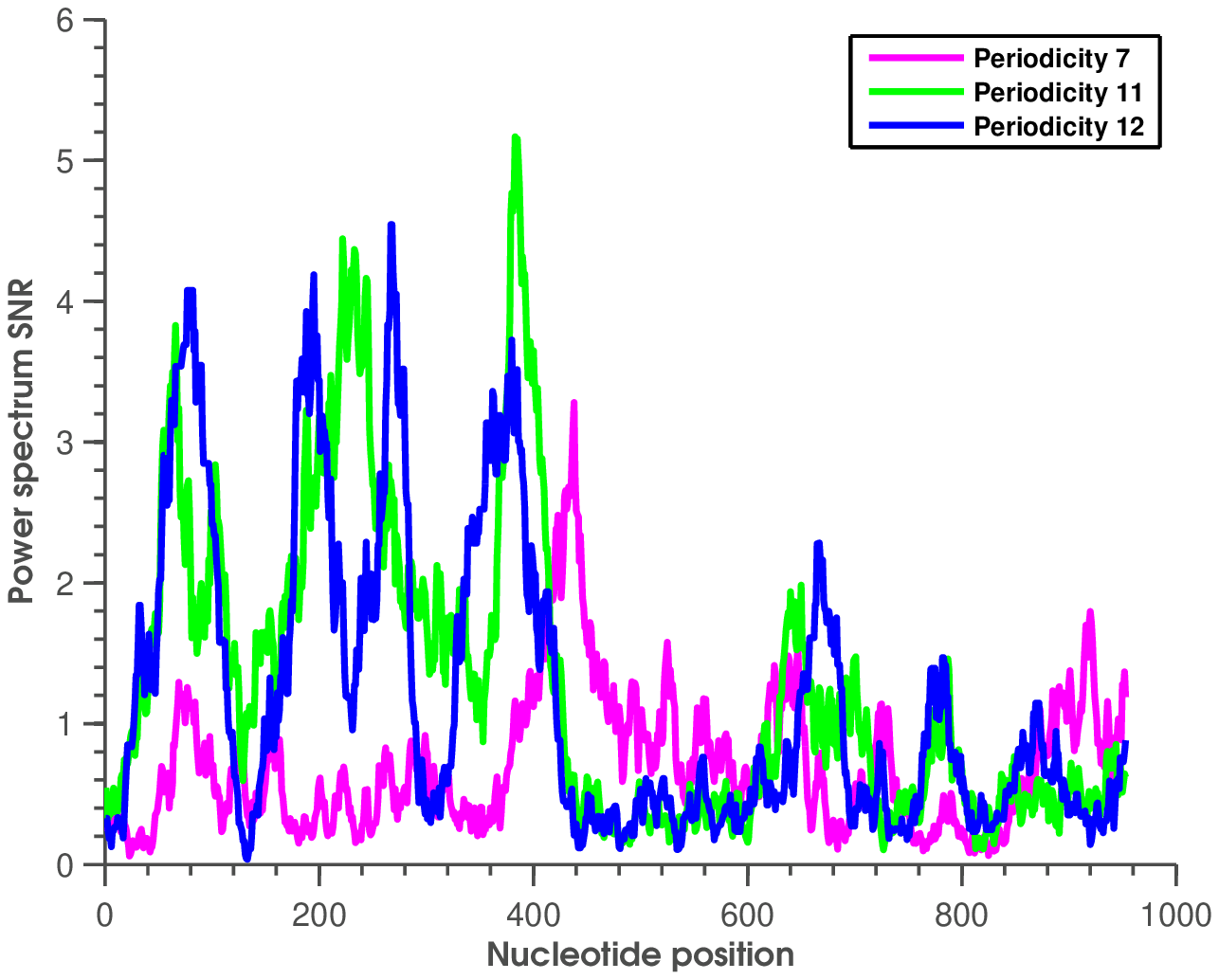}}\quad
       \caption{Power spectra of DNA sequence M65145. (a) Fourier power spectrum, (b) PPS spectrum SNR, (c) sliding window PPS spectrum SNRs of different periodicities, window size = 60 bp.}
       \label{fig:sub1}
     \end{figure}
 %Figure 5
 The second example is periodicity detection in Human microsatellite repeats (GenBank locus:HSVDJSAT, 1985 bp) using the PPS method. This DNA sequence contains variable length tandem repeats (VLTRs) \citep{hauth2002beyond}. Figure 5(a) is the Fourier power spectrum of the DNA sequence and suggests. It indicates the complexity and high noise level in Fourier power spectrum, and thus it is difficult to detect periodicities from Fourier power spectrum. From the PPS spectrum in Figure 5(b), we can detect accurately the short periodicities: 4, 6, 8, 10, 22, 49, and 50. The corresponding SNR values of these periodicities are: $P_4$:1.2781, $P_6$:1.4620, $P_8$:1.0349, $P_{10}$:1.7631, $P_{22}$:1.0781,$P_{49}$:1.1268,$P_{50}$:1.1023. The strongest periodicity in this DNA sequence is 10 periodicity, which is in agreement with literature methods \citep{hauth2002beyond,gupta2007novel}. From the sliding window PPS spectra of periodicities 4, 6 and 10 in Figure 5(c), we can locate the positions of these periodicities. For example, a major location of the periodicity 4 is between 700-900 bp; the periodicities 6 is mainly located around 400 bp and 1000 bp, periodicities 10 is mainly located at 1180 bp and 1400 bp. The previously studies \citep{hauth2002beyond,gupta2007novel} can identify similar periodicities but those methods need different threshold settings and sometimes get conflicting results.
 
%File:DNA_Profile/DNAProfile_HSVDJSAT_01082015
  \begin{figure}[tbp]
         \centering
         \subfloat[]{\includegraphics[width=3.25in]{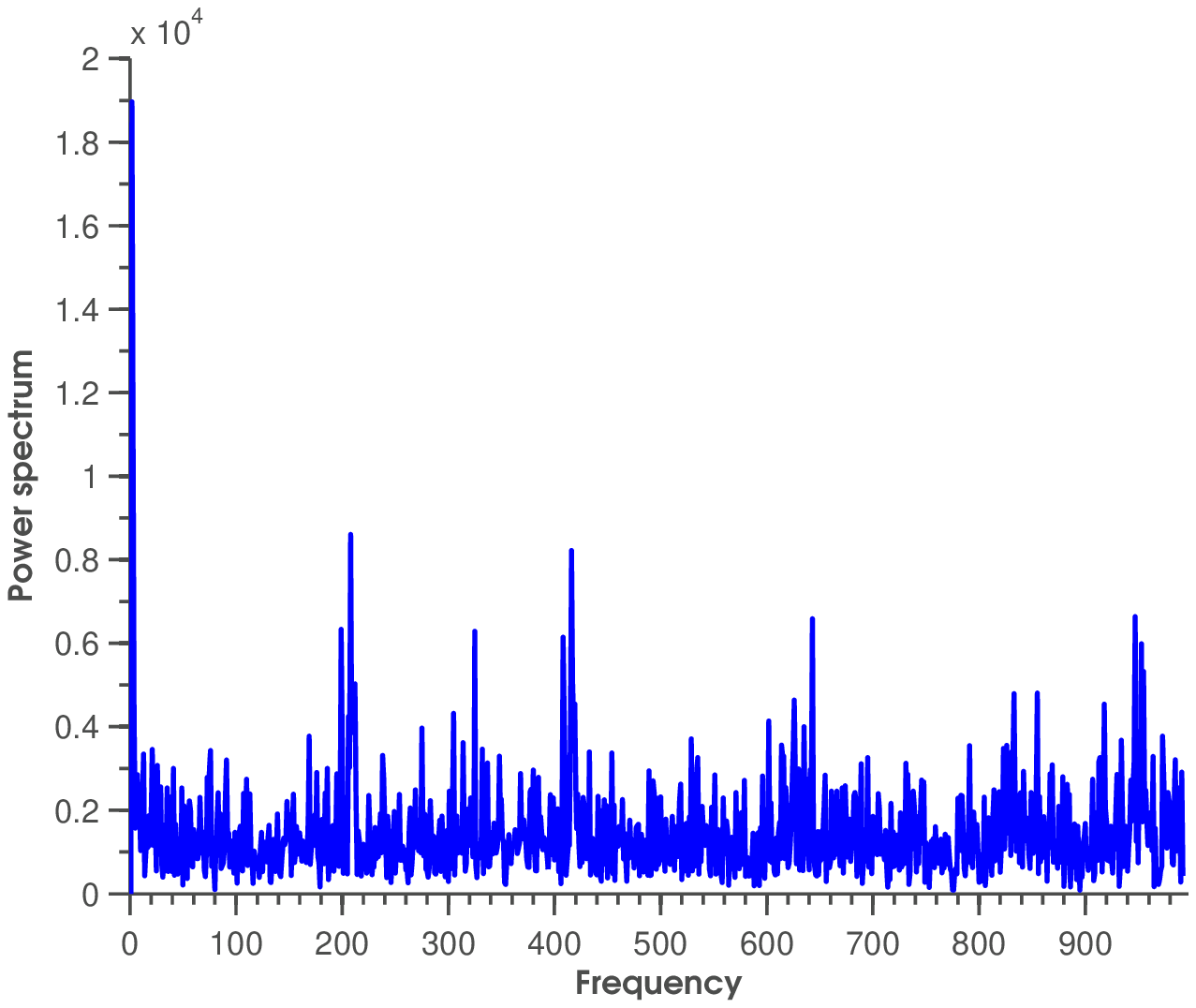}}\quad
         \subfloat[]{\includegraphics[width=3.25in]{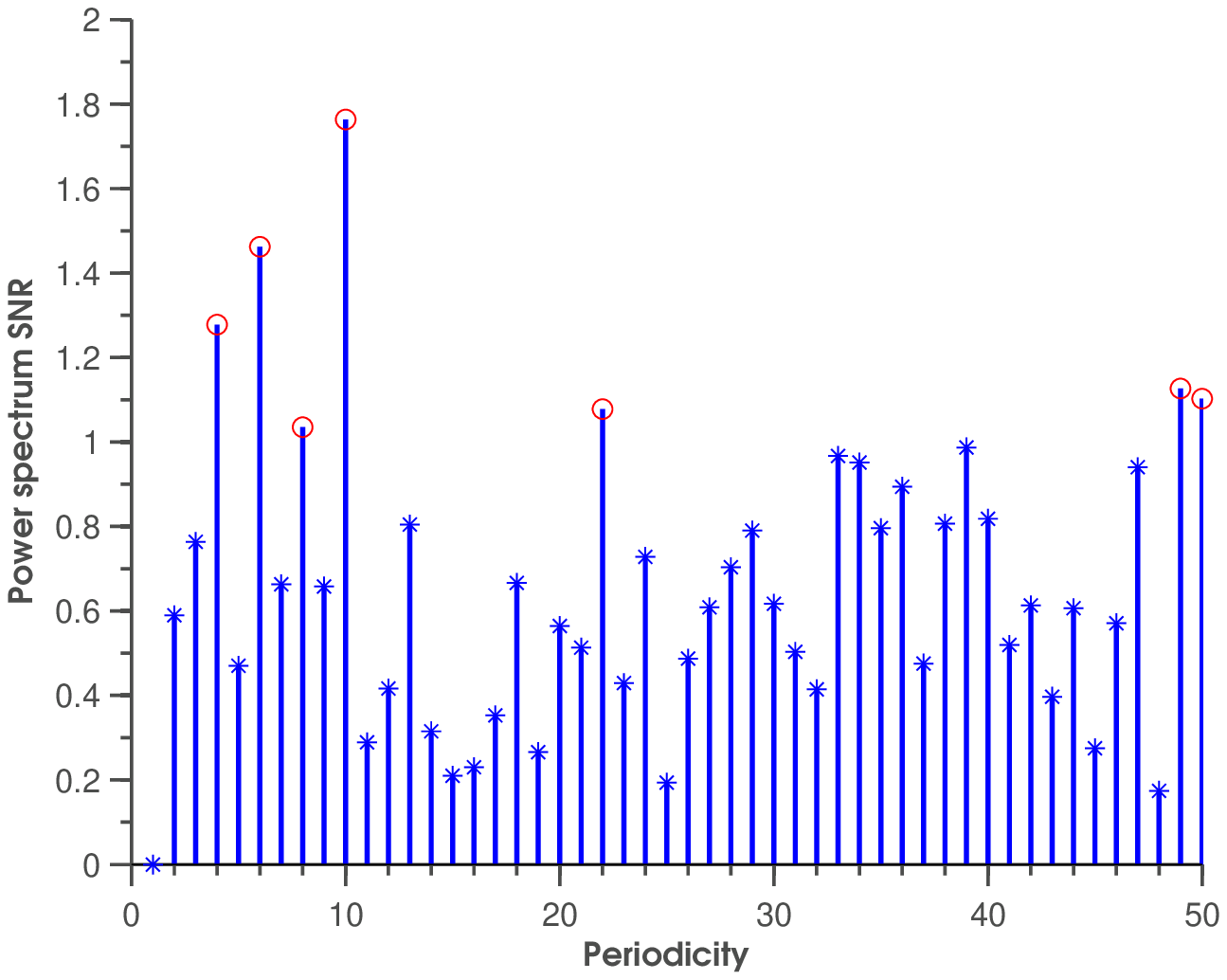}}\quad
         \subfloat[]{\includegraphics[width=3.25in]{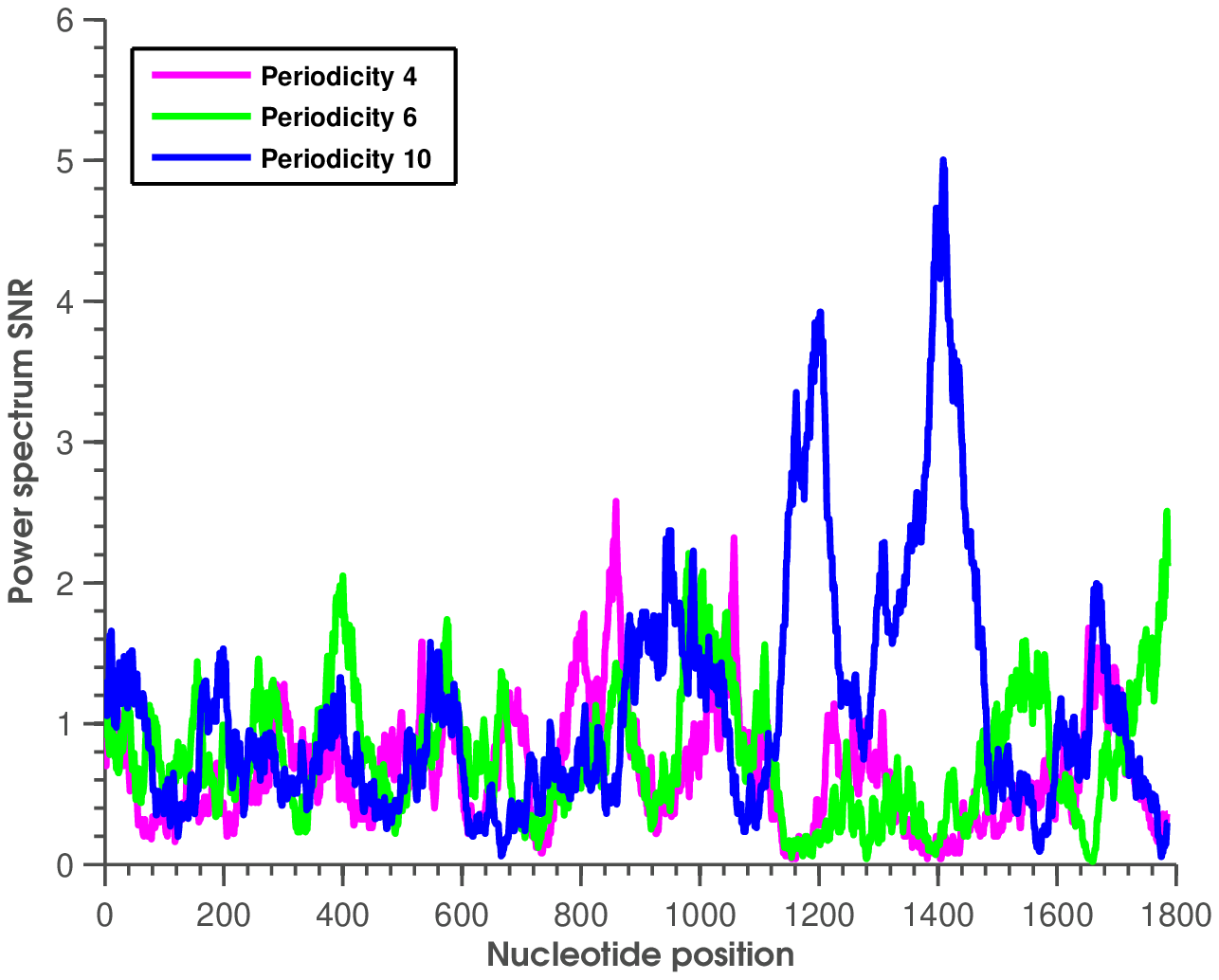}}\quad
         \caption{Power spectra of DNA sequence HSVDJSAT. (a) Fourier power spectrum, (b) PPS spectrum SNR, (c) sliding window PPS spectrum SNRs of different periodicities, window size = 100 bp.}
         \label{fig:sub1}
       \end{figure}
       
These examples demonstrate that the PPS spectrum can detect a broad range of latent and complex periodicities in DNA sequences, which are missed in current state of art methods.
 \begin{comment}
 \subsection{Periodicities in protein coding regions and LncRNA genes}
 %Figure 6 
 Recent advances in long noncoding RNAs (lncRNAs) uncovered show that lncRNAs become recognized as major players for epigenetic regulation. For example, MicroRNA-23a promotes myelination in the central nervous system \citep{lin2013microrna}. We here study the periodicities within the lncRNA sequence MicroRNA-23a. We studied the short periodicities in protein coding regions, introns, and non-coding RNA genes (LncRNA). Figure 6(a) PPS power spectrum SNR means of protein coding regions at different short periodicities. It shows that protein coding region has strong periodicity 3 and weak periodicities from 2 to 12. Figure 6(b) PPS  spectrum SNR means for LncRNA at different short periodicities. It shows that short periodicities including periodicities  2 to 12 are all weak. But for PPS spectra of intron sequences, the mean values for all these periodicities are zeros (data not shown).
   
   %File:DNA_Profile/DNAProfile_EILNC_01252015.m
     \begin{figure}[tbp]
         \centering
         \subfloat[]{\includegraphics[width=3.25in]{DNAProfile_EILNC_01252015_fig1.eps}}\quad
         \subfloat[]{\includegraphics[width=3.25in]{DNAProfile_EILNC_01252015_fig2.eps}}\quad
         \caption{PPS power spectrum SNR for (a) protein coding regions, (b) LncRNA genes.}
         \label{fig:sub1}
       \end{figure}
\end{comment}
\subsection{Computational complexity}
One of the challenging problems when applying Fourier transform to sequence study is the computational complexity. Fourier transform attempts to explain a data set as a weighted sum of sinusoidal basis element when the data can be closely approximated by such elements, however, directly searches for periodicities, repetitions, or regularities in the data. However, Fourier transform is computational intensive technique. The FFT still needs ${\rm O}(N\log N)$ computational time. Because our PPS algorithm only computes the spectrum for periodicities of special interests, there is no need to compute Fourier transform for all frequencies, it can significantly reduce the computation complexity. In addition, since the spectrum transform matrix $S_p$ is real number matrix, computation of PPS is performed only on real numbers, not complex numbers, this reduces computation time of the PPS algorithm.

\subsection{Discussion}
From this study, the major advantage of the PPS spectrum is that it can capture all the tandem periodicities because the PPS spectrum reduces high noise level and spectral leakage as in Fourier spectrum. It is worthy noting that the power spectrum method in this paper can also be applied on detection periodicities in other time series such as protein sequence and biomedical signals. An open question is that the PPS spectrum after a specific periodicity becomes smooth and lacks of characters of peaks (Figure 1(c), 2(c), 3(b)). We examine the value this specific periodicity, we found the position approximately equals to $\sqrt {2n}$. We will investigate this problem in future studies. 

\section{Conclusions} 
Periodic latent regions in genomes often play important roles in many genomic functions. Carefully understanding the source and pattern of periodicity in genomic sequences thus represents an important problems in biology. 
 
In this paper, we propose a novel power spectrum based on the periodic distribution of symbols. We also propose an algorithm to compute PPS spectrum. The algorithm can identify and locate exact and inexact periodic patterns in DNA sequences. The algorithm has been assessed for identification of periodicities in different genomic elements, including exons and microsatellite DNA sequences. The results. The experimental results demonstrate the effectiveness of our algorithm on periodicity identification and minimizing spectral leakage. The results demonstrate the utility of analyzing the genomes in the periodicity space and our method can be deterministic method for periodicity identification.

%%%%%%%%%%%%%%%%%%%%%%%%%%%%%%%%%%%%%%%%%%%%%%%%%%%%%%%%%%%%%%%%%%%%%%%%%%%%%%%%%%%%%
%
%     please remove the " % " symbol from \centerline{\includegraphics{fig01.eps}}
%     as it may ignore the figures.
%
%%%%%%%%%%%%%%%%%%%%%%%%%%%%%%%%%%%%%%%%%%%%%%%%%%%%%%%%%%%%%%%%%%%%%%%%%%%%%%%%%%%%%%
%\section*{Acknowledgement}
%We thank Andrew Yin for proof reading of the manuscript.
%\paragraph{Funding\textcolon} 

%\bibliographystyle{natbib}
%\bibliographystyle{achemnat}
%\bibliographystyle{plainnat}
%\bibliographystyle{abbrv}
%\bibliographystyle{bioinformatics}
%
%\bibliographystyle{plain}
%
%\bibliography{Document}

\begin{comment}

\end{comment}

\bibliographystyle{elsarticle-harv}
%\bibliographystyle{elsarticle-num}
%\bibliography{jtbRefs}
\bibliography{../References/myRefs}
\appendix
\renewcommand\thefigure{\thesection.\arabic{figure}}
\setcounter{theorem}{0}
\setcounter{figure}{0}
\setcounter{table}{0}
\section*{Supplementary Materials}
 \subsection*{The spectrum transform matrices (Table 1) and DNA sequences}
 N130P5=CCATATCCGATCGGCAGCGCGTGCC
    TTTTATCGCTATCGATCGAATGGGCTCGAGGA
    CCGCGGCTGTCTATAGAAAAATTATAAATGAT
    ATTGATCCGAGTAGGGTCCCACTCGGTGCGGG
    GCACTTCAA\\
    
{\normalsize {\small {\normalsize }}}N130P5-D2=CCATATCCGATCGGCAGCGCG
    TGCCTTTTATCGCTATCGATCGAATGGGCTC
    GAGGACCGCGGCTGTCTATAGAAAAATTATA
    AATGATATTGATCCGAGTAGGGTCCCACTCG
    GTGCGGGGCACTTC
    
  %\subsection*{Spectrum transform matrices}
      \begin{table}[tbp]
      \caption{Spectrum transform matrices of some short periodicities} % title of Table
      \centering % used for centering table
      \begin{tabular}{c|c} % centered columns (4 columns)
      \hline\hline %inserts double horizontal lines
      Periodicity  & Spectrum transform matrix $S_p$  \\ [0.45ex] % inserts table
      %heading
       \hline
       %  inserts single horizontal line
       \\[0.005in]
       $P_{2}$  & $\begin{bmatrix}
              1 &    0\\
                 -2  &   1\\
          \end{bmatrix}$ \\ 
           \\[0.005in] \hline  \\[0.005in] %  inserts single horizontal line
        $P_{3}$  & $\begin{bmatrix}
             1    &     0   &      0\\
             -1  &  1   &      0\\
            -1 &  -1 &   1\\
         \end{bmatrix}$ \\ 
            \\[0.005in] \hline \\[0.005in] 
        $P_{4}$  & $\begin{bmatrix}
             1  &       0  &       0   &      0\\
             0  &  1  &       0   &      0\\
            -2  &  0  &  1   &      0\\
            0  & -2  &  0   & 1\\
         \end{bmatrix}$ \\ 
             \\[0.005in] \hline \\[0.005in] 
        $P_{5}$  & $\begin{bmatrix}
             1&         0&         0 &        0   &      0\\
             \frac{\sqrt 5  - 1}
             {2}&    1&         0 &        0   &      0\\
            - \frac{\sqrt 5  + 1}
            {2}&    \frac{\sqrt 5  - 1}
            {2}&    1 &        0   &      0\\
            - \frac{\sqrt 5  + 1}
            {2}&   - \frac{\sqrt 5  + 1}
            {2}&    \frac{\sqrt 5  - 1}
            {2} &   1   &      0\\
            \frac{\sqrt 5  - 1}
            {2}&  - \frac{\sqrt 5  + 1}
            {2}&   - \frac{\sqrt 5  + 1}
            {2} &   \frac{\sqrt 5  - 1}
             {2}   & 1\\
         \end{bmatrix}$ \\ 
             \\[0.005in] \hline \\[0.005in] 
        $P_{6}$  & $\begin{bmatrix}
               1 &        0&         0 &        0 &        0   &      0\\
               1 &   1&         0 &        0 &        0   &      0\\
              -1 &   1&    1 &        0 &        0   &      0\\
              -2 &  -1&   1  &  1  &       0    &     0\\
              -1 &  -2 &  -1 &   1 &   1   &      0\\
               1 &  -1 &  -2 &  -1 &   1   & 1\\
           \end{bmatrix}$ \\
            \\[0.005in] 
           \hline\hline %inserts single line
      \end{tabular}
      \label{table:nonlin} % is used to refer this table in the text
      \end{table}
       
\begin{comment}
 \begin{prop}
   For a DNA sequence of length $n$, the nucleotide distribution of the sequence can be represented by the PPS power spectrum at periodicities $1, 2, \cdots , p$, where the upper bound of $p$ for the periodicities is as follows
   \begin{equation}
   % MathType!MTEF!2!1!+-
   % feaafiart1ev1aaatCvAUfeBSjuyZL2yd9gzLbvyNv2CaerbuLwBLn
   % hiov2DGi1BTfMBaeXatLxBI9gBaerbd9wDYLwzYbItLDharqqtubsr
   % 4rNCHbWexLMBbXgBd9gzLbvyNv2CaeHbl7mZLdGeaGqiVCI8FfYJH8
   % YrFfeuY-Hhbbf9v8qqaqFr0xc9pk0xbba9q8WqFfeaY-biLkVcLq-J
   % Hqpepeea0-as0Fb9pgeaYRXxe9vr0-vr0-vqpWqaaeaabiGaciaaca
   % qabeaadaqaaqaafaGcbaGaamiCaiabg2da9maahmaabaWaaSaaaeaa
   % cqGHsislcaaIXaGaey4kaSYaaOaaaeaacaaI4aGaamOBaiabgUcaRi
   % aaigdaaSqabaaakeaacaaIYaaaaaGaayP74laawMp+aaaa!4CEA!
  % \[
   p = \left\lceil {\frac{{ - 1 + \sqrt {8n + 1} }}
   {2}} \right\rceil 
   %\]. 
   \end{equation}
   , where $\left\lceil {} \right\rceil $ is the ceiling function of a real number. The upper bound $p$ can be furthered approximated as  $p = \left\lceil {\sqrt {2n} } \right\rceil$.
   \end{prop}
   
   Justification: PPS power spectrum at periodicity $p$ corresponds to the nucleotide distribution $Fp$. 
   \end{comment}

\end{document}